\documentclass[sigconf,authorversion,nonacm]{acmart}

\usepackage{xspace}
\usepackage{graphicx}
\usepackage{xcolor}
\usepackage{caption}
\usepackage{subcaption}
\usepackage{balance}
\usepackage{booktabs}
\usepackage{tabularx}
\usepackage{diagbox}
\usepackage[utf8]{inputenc}
\usepackage{makecell}
\usepackage{hyperref}
\usepackage{enumitem}
\usepackage{nicematrix}
\usepackage{tikz}
\usepackage{pifont}
\usepackage{graphics}
\usepackage{multirow}

\usepackage[linesnumbered,ruled,vlined]{algorithm2e}
\usepackage{setspace}

\usepackage{cleveref}
\newcommand{\lst}{\textsc{LST}\xspace}
\newcommand{\lsts}{\textsc{LSTs}\xspace}
\newcommand{\framework}{\textsc{AutoComp}\xspace}

\newcommand{\cut}[1]{}

\newcommand{\smallsection}[1]{\vspace{1mm}\noindent\textbf{#1.}}	%
\newcommand{\newparagraph}{\vspace{0.5mm}\noindent}

\usetikzlibrary{patterns, shadings}

\newcommand{\myparagraph}[1]{\vspace{1mm}\noindent\textbf{#1}}

\newcommand{\functional}[1]{\textbf{$\textrm{FR}$#1}\xspace}
\newcommand{\nonfunctional}[1]{\textbf{$\textrm{NFR}$#1}\xspace}

\begin{document}

\title{\framework: Automated Data Compaction for\\Log-Structured Tables in Data Lakes}
\renewcommand{\shorttitle}{\framework: Automated Data Compaction for Log-Structured Tables in Data Lakes}

\settopmatter{authorsperrow=4}
\author{Anja Gruenheid}
\authornote{Authors contributed equally.}
\affiliation{%
  \institution{Microsoft}
  \city{Zurich}
  \country{Switzerland}
}
\author{Jes\'us Camacho-Rodr\'iguez}
\authornotemark[1]
\affiliation{%
  \institution{Microsoft}
  \city{Mountain View}
  \state{CA}
  \country{USA}
}
\author{Carlo Curino}
\affiliation{%
  \institution{Microsoft}
  \city{Redmond}
  \state{WA}
  \country{USA}
}
\author{Raghu Ramakrishnan}
\affiliation{%
  \institution{Microsoft}
  \city{Redmond}
  \state{WA}
  \country{USA}
}
\author{Stanislav Pak}
\authornotemark[1]
\affiliation{%
  \institution{LinkedIn}
  \city{Sunnyvale}
  \state{CA}
  \country{USA}
}
\author{Sumedh Sakdeo}
\authornotemark[1]
\affiliation{%
  \institution{LinkedIn}
  \city{Sunnyvale}
  \state{CA}
  \country{USA}
}
\author{Lenisha Gandhi}
\affiliation{%
  \institution{LinkedIn}
  \city{Sunnyvale}
  \state{CA}
  \country{USA}
}
\author{Sandeep K. Singhal}
\affiliation{%
  \institution{LinkedIn}
  \city{Sunnyvale}
  \state{CA}
  \country{USA}
}
\author{Pooja Nilangekar}
\authornote{Work done while author was at Microsoft.}
\affiliation{%
  \institution{University of Maryland}
  \city{College Park}
  \state{MD}
  \country{USA}
}
\author{Daniel J. Abadi}
\affiliation{%
  \institution{University of Maryland}
  \city{College Park}
  \state{MD}
  \country{USA}
}
\renewcommand{\shortauthors}{Anja Gruenheid et al.}



\begin{abstract}
The proliferation of small files in data lakes poses significant challenges, including degraded query performance, increased storage costs, and scalability bottlenecks in distributed storage systems. 
Log-structured table formats (\lsts) such as Delta Lake, Apache Iceberg, and Apache Hudi exacerbate this issue due to their append-only write patterns and metadata-intensive operations. 
While compaction--the process of consolidating small files into fewer, larger files--is a common solution, existing automation mechanisms often lack the flexibility and scalability to adapt to diverse workloads and system requirements while balancing the trade-offs between compaction benefits and costs.
In this paper, we present \framework, a scalable framework for automatic data compaction tailored to the needs of modern data lakes. 
Drawing on deployment experience at LinkedIn, we analyze the operational impact of small file proliferation, establish key requirements for effective automatic compaction, and demonstrate how \framework addresses these challenges. 
Our evaluation, conducted using synthetic benchmarks and production environments via integration with OpenHouse--a control plane for catalog management, schema governance, and data services--shows significant improvements in file count reduction and query performance. 
We believe \framework's built-in extensibility provides a robust foundation for evolving compaction systems, facilitating future integration of refined multi-objective optimization approaches, workload-aware compaction strategies, and expanded support for broader data layout optimizations. 
\end{abstract}

\maketitle

\section{Introduction}
In recent years, enterprises have undergone a significant shift in their approach to data management, progressively gravitating towards data lake-centric architectures. 
Data lakes originated as cost-effective storage for large volumes of unstructured, uncleaned, or ungoverned data in scalable distributed file systems like HDFS, providing an alternative to storing this data in expensive proprietary data management or file systems. 
Over time, the declining cost of data lake storage has encouraged organizations to use these systems for managing core, governed, and structured data as well. 
This shift was enabled by the widespread adoption of scalable storage services~\cite{aws-s3,azure-adls,google-cloud-storage,ozone} and efficient open-source data formats~\cite{parquet, orc} that serve as foundational elements for persisting data across diverse workloads.
Data stored in distributed storage systems is then accessible to various engines and applications, providing several advantages:
($i$)~independent scaling of storage and compute, enhancing efficiency and cost savings, 
($ii$)~elimination of data silos, which streamlines workflows and simplifies complex data movement across systems, and 
($iii$)~the flexibility to choose the optimal engine for each application, thereby mitigating lock-in concerns. 
Various commercial platforms embrace this approach~\cite{databricks,fabric,snowflake}.

Engines and applications accessing these distributed storage systems require guarantees such as consistency and isolation during complex transactions involving read and write operations.
However, these storage systems are primarily designed for scalability and durability, and lack the concurrency and recovery capabilities needed to meet these requirements. 
As a result, open table formats such as Delta Lake~\cite{delta-lake,DBLP:journals/pvldb/ArmbrustDPXZ0YM20}, Apache Iceberg~\cite{apache-iceberg}, and Apache Hudi~\cite{apache-hudi}, also referred to as log-structured tables or \lsts in the following, have emerged to enable structured data to be stored in data lake storage solutions while remaining organized and optimized for external query engines, achieving excellent query performance.

At their core, these \lsts store data persistently in immutable files relying on open-source columnar formats~\cite{parquet,orc} and propose ($i$)~a metadata layer that records table versions and attributes such as data schemas and statistics, and ($ii$)~a protocol to coordinate interactions with a table during read and write operations. 
Catalogs play a critical role in this context by maintaining references to table metadata and enabling seamless access and updates across various systems~\cite{openhouse,apachepolaris,unity}. 
With each write operation, new data files are added to the table, and the corresponding table metadata is updated. 
Over time, layers of data files (small in size in common trickle-write scenarios and untuned writers) can accumulate within the table.

\smallsection{The Challenge of Small Files} 
The accumulation of numerous small files presents a significant challenge in data lake-centric architectures, impacting all engines and \lst implementations, as extensively documented in prior studies~\cite{2024lstbench,delta-1gb,aws-iceberg-optimize,dremio-small-data-files}. 
This proliferation of small files increases overhead due to a higher number of managed objects and more frequent IO requests, which can strain the distributed storage systems underpinning data lakes and impact their performance and scalability~\cite{DBLP:conf/sigmod/RamakrishnanSDK17,linkedin-hdfs}. 
For instance, HDFS encounters scalability challenges as the NameNode, which maintains file system metadata, can manage only a limited number of objects. 
As file counts grow, the number of managed objects rises proportionally, placing additional pressure on the NameNode and often necessitating federation to distribute the load. 
Additionally, elevated RPC traffic generated by small files places further burden on HDFS, requiring additional observer NameNodes (i.e., read-only replicas) to manage the increased traffic effectively.

Small files storing a limited number of rows also reduce the efficiency of columnar formats that rely on robust encoding and compression to optimize data access and storage. 
Moreover, the presence of these files contributes to bloated metadata in \lsts. 
Each transaction appends references to the files in logs or manifests, causing metadata size to grow and increasing the time required for query processing and maintenance operations, thereby affecting overall performance and efficiency. 
The problem is exacerbated by potential monetary costs, as cloud service providers often charge based on IO requests and data transfer volume~\cite{aws-price-calc,azure-price-calc,gcp-price-calc}.

\smallsection{Compaction as a Solution} 
The most prevalent \emph{storage healing} mechanism to address this issue is \emph{compaction}.
Compaction is the process of rewriting data files in a table to create fewer, larger files according to a target file size, which helps improve storage efficiency, query performance (both in planning and execution), and overall data organization. 
Each \lst implements its own version of compaction mechanisms~\cite{delta-compaction,iceberg-compaction,hudi-compaction}. 
Industry practices for compaction vary widely~\cite{synapse-optimized-write, databricks-predictiveopt, aws-iceberg-optimize, netflix-auto-optimize}, from reactive strategies that trigger compaction after a data write operation to standalone solutions that periodically optimize the storage layout. 
While some engines integrate proactive mechanisms to maintain an optimal data layout, these optimizations are typically performed in isolation, addressing only the needs of the specific engine without considering other engines that might access the same data.

\smallsection{Contributions}
In this work, we address the challenge of small file proliferation in \lsts by drawing on our practical experience to develop an automated data compaction solution. 
Our contributions are as follows:

\begin{itemize}[leftmargin=*]
\item \textbf{Analysis of Small File Proliferation in Industry Scenario.} 
We highlight the operational challenges posed by data fragmentation across numerous small files in a real-world scenario at LinkedIn. 
We identify the most common causes of this small file proliferation and illustrate its impact on storage efficiency and query performance (\S\ref{sec:motivation}).

\item \textbf{Definition of Requirements and Introduction of \framework.} 
Guided by our findings, we establish a set of essential functional and non-functional requirements to effectively address small file proliferation in \lsts in practice. 
We then introduce \framework, a framework designed to meet these requirements and enable automatic scalable data compaction (\S\ref{sec:overview}-\ref{sec:auto-datacompact}).

\item \textbf{Comprehensive Evaluation of \framework.} 
We evaluate \framework in a cloud-based deployment using synthetic benchmarks to assess its effectiveness. 
We also report on its impact after deployment at LinkedIn, demonstrating substantial improvements of up to 44\% reduction in the number of files smaller than ``128MB'' in a production environment (\S\ref{sec:synt_eval}-\ref{sec:impactpractice}).

\item \textbf{Discussion of Future Directions.} 
We identify areas for improvement and propose future research directions to improve \framework and data compaction techniques in \lsts (\S\ref{sec:future}).
\end{itemize}

\section{Motivating Scenario}
\label{sec:motivation}

At LinkedIn, raw event data is ingested from thousands of services into a data lake through a centrally managed pipeline powered by Apache Gobblin~\cite{linkedin-gobblin}. 
The organization is structured into various lines of business, each responsible for maintaining data pipelines that produce derived data, business metrics, and feature sets. 
Line-of-business engineers, referred to as end-users in the following to distinguish them from data infrastructure engineers, primarily develop these pipelines using compute engines such as Apache Spark, Trino, and Apache Flink.

\begin{figure}[t]
    \centering
    \includegraphics[width=.85\columnwidth]{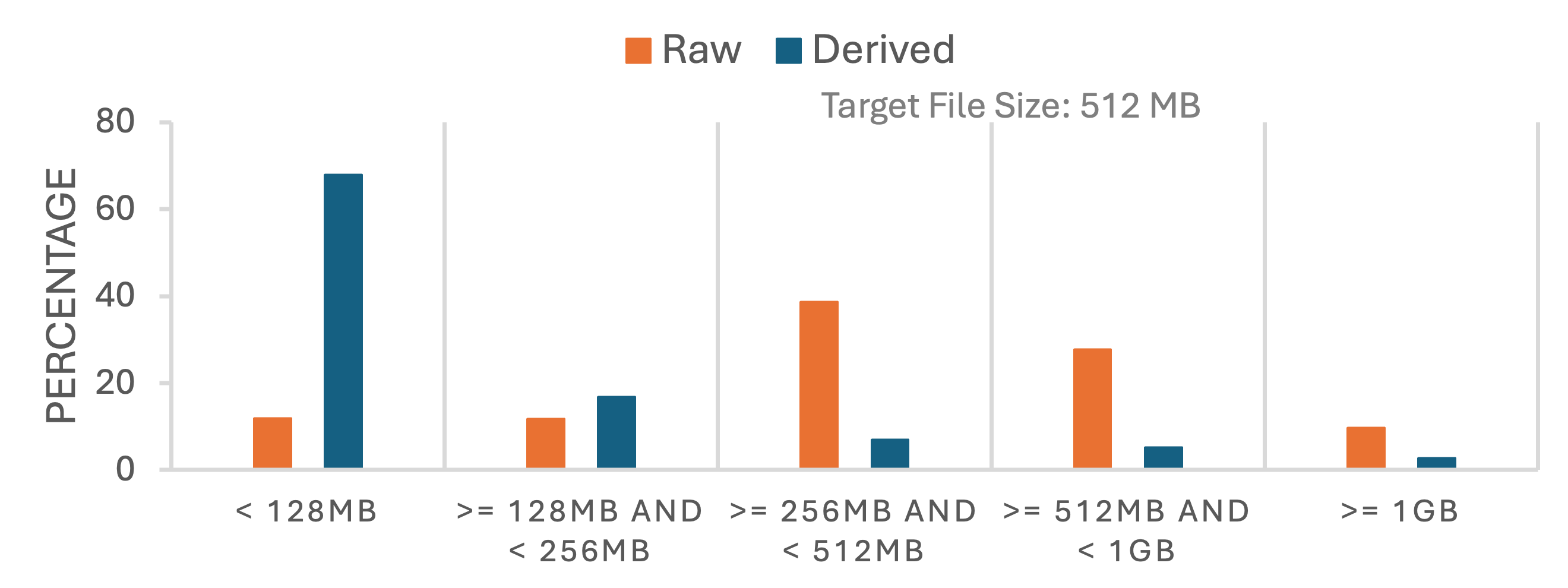}
    \vspace{-0.5em}\caption{File size distribution for ingested data (raw ingestion vs. user-derived data).}\vspace{-0.5em}
    \label{fig:back:raw_derived_dist}
\end{figure}

LinkedIn has adopted Apache Iceberg as the \lst for storing data generated by these pipelines, thereby standardizing data storage practices across its analytics and artificial intelligence workloads. 
Building on the adoption of Iceberg, LinkedIn has also developed and open-sourced \textbf{OpenHouse}~\cite{openhouse}, 
a control plane that provides a declarative catalog for table definitions, schema management, and metadata maintenance, along with data services to reconcile observed and desired states. 
For over a year, LinkedIn has been onboarding existing and new tables into OpenHouse.
The distribution of file sizes exhibits a marked difference between raw data ingested by the central pipeline and derived data generated by end-user jobs, as illustrated in \Cref{fig:back:raw_derived_dist}. 
The central pipeline follows a well-defined pattern, writing raw event data from Kafka to HDFS every five minutes and incrementally compacting and deduplicating it into hourly partitions, resulting in files of approximately $512$MB in size, i.e.,~our target file size. 
Daily partitions, composed of $24$~hourly segments, are retained for long-term storage, while smaller checkpoint files are expired after three days. 
In contrast, end-user jobs using Spark, Trino, and Flink are neither designed nor tuned for generating optimal file sizes, resulting in a high concentration of small files. 
Expecting end-users to prioritize file optimization is unrealistic, as their primary focus is addressing business challenges rather than managing low-level data storage concerns.

\smallsection{Causes of Small File Existence} 
The proliferation of small files in these derived tables can be attributed to several factors related to Iceberg's versioning semantics as well as the design and configuration of end-user jobs. 
($i$)~\emph{Inserts}: 
While bulk inserts can produce optimally sized files, aspects such as engine configuration, degree of parallelism, and memory constraints significantly influence the resulting number of files in the table. 
Incremental inserts, including Change Data Capture (CDC) scenarios~\cite{iceberg-cdc}, often lead to the rapid creation of numerous small files. 
($ii$)~\emph{Updates and Deletes}: 
In Copy-on-Write (CoW) configurations, deletions can affect data distribution across files, leading to uneven file sizes.
Merge-on-Read (MoR) configurations generate delta files that accumulate over time. 
($iii$)~\emph{Migration}: 
When existing Parquet or ORC data is migrated into Iceberg, the original file structure is typically preserved while Iceberg metadata is layered on top~\cite{iceberg-migrate}, resulting in suboptimal file layouts. 
($iv$)~\emph{Metadata}: 
Iceberg introduces additional metadata for each table to manage state, including manifests and manifest lists. 
This added metadata contributes to small file proliferation.

\begin{figure}[t]
    \centering
    \includegraphics[width=\columnwidth]{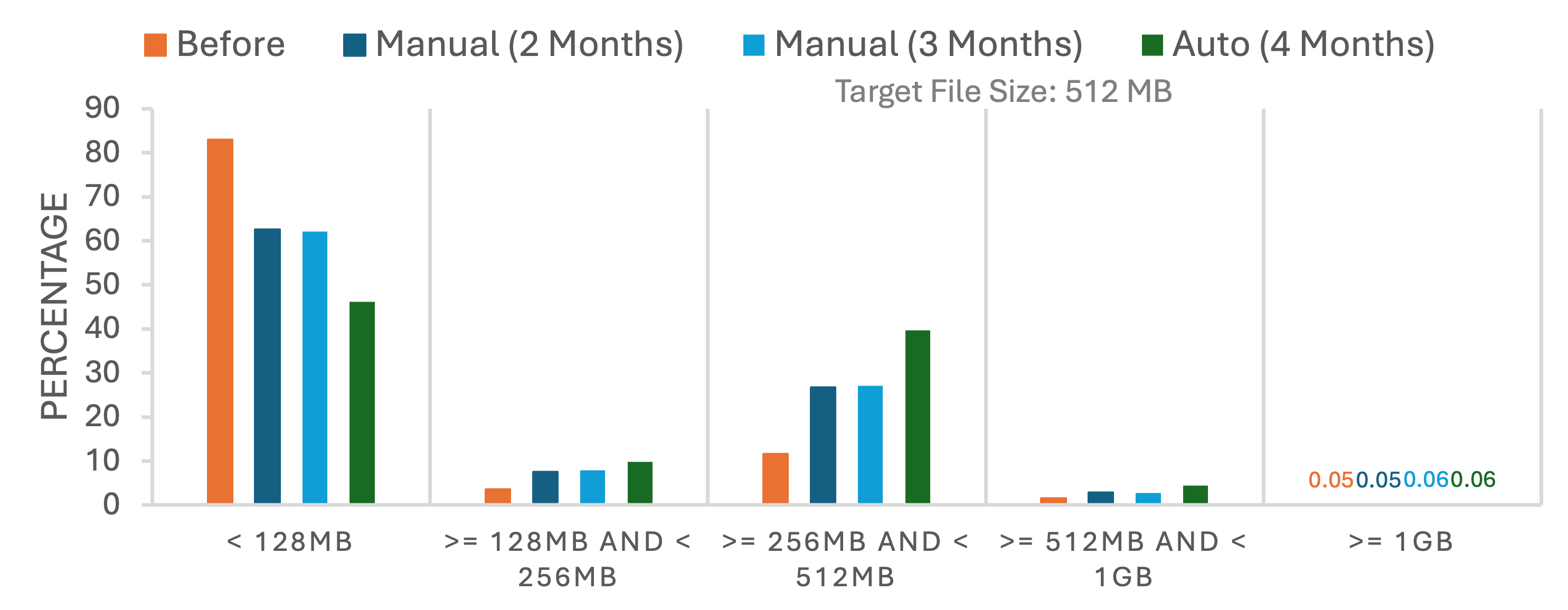}
    \vspace{-2em}\caption{File size distribution for OpenHouse-managed Iceberg tables, shown before and after compaction.}
    \label{fig:back:oh-distribution}
\end{figure}

\begin{figure}[t]
    \centering
    \includegraphics[width=.75\columnwidth]{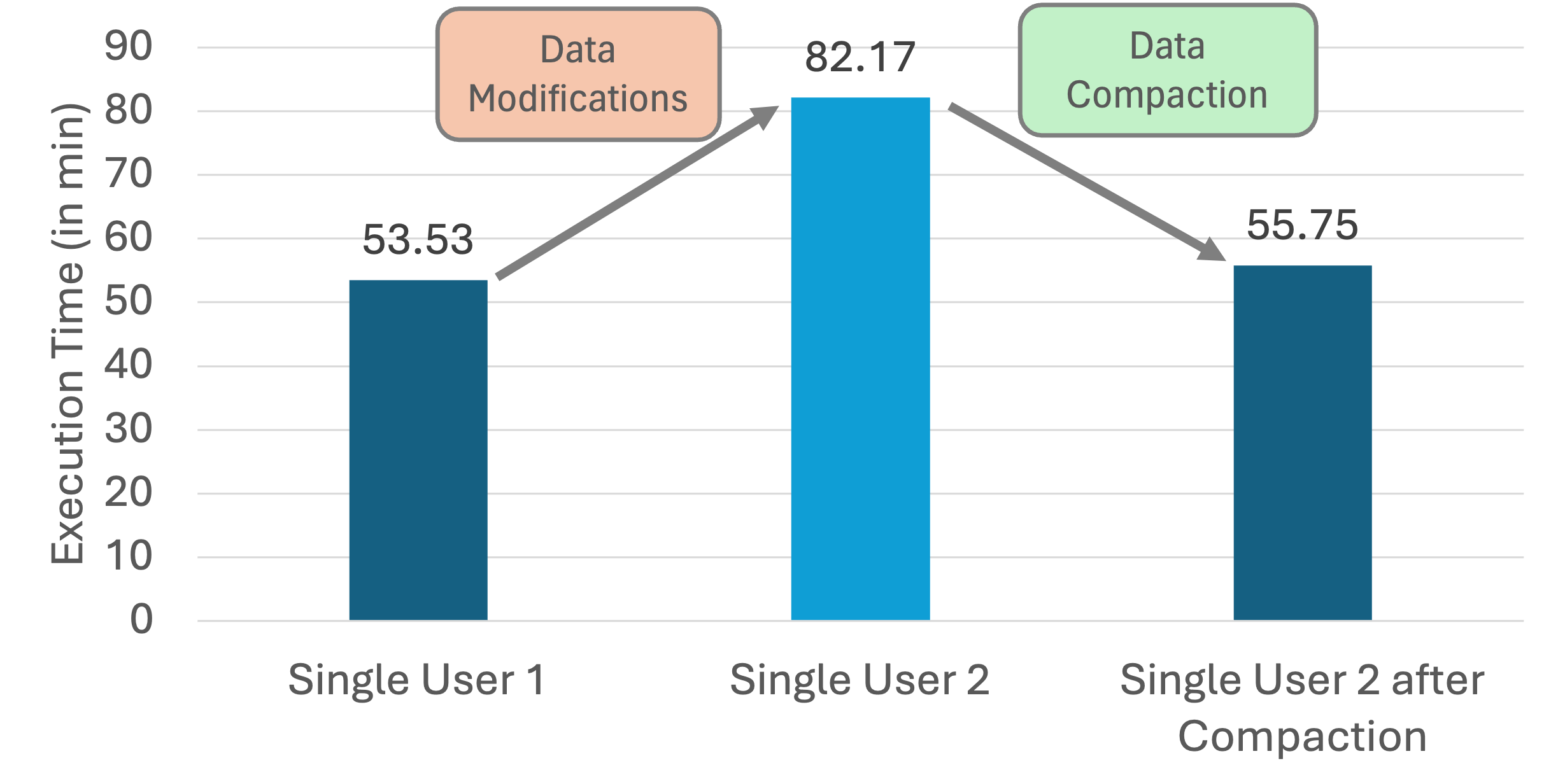}
    \vspace{-1.25em}\caption{TPC-DS experiment (Apache Spark \& Iceberg): Comparison of execution time before and after compaction.}\vspace{-0.5em}
    \label{fig:back:fragmentation}
\end{figure}

\smallsection{Impact of Compaction}
The small file size distribution patterns observed in user-derived data prompted data infrastructure engineers to leverage OpenHouse to introduce centrally managed compaction for the first time at LinkedIn.
In the initial implementation, compaction was triggered manually for selected tables that exhibited recurring issues such as query failures, quota breaches, and namespace growth in HDFS.
This approach proved effective; as shown in~\Cref{fig:back:oh-distribution}, manual compaction realigned file size distribution toward the target, reducing the storage system's load and improving overall efficiency.

In addition to its impact on the storage layer, we also note that file proliferation impacts query performance. 
However, directly measuring the impact of compaction on query performance is challenging in a production environment, as infrastructure engineers do not control workloads executed by different lines of business. 
To provide insights, we conducted a synthetic experiment using the TPC-DS benchmark~\cite{DBLP:conf/vldb/OthayothP06} at a scale factor of 1000. 
The results, shown in \Cref{fig:back:fragmentation}, capture the end-to-end runtime of the \emph{single-user} phase, which includes all TPC-DS queries, on a 16-node Spark cluster before and after a \emph{data maintenance} phase. 
During the \emph{data maintenance} phase, about 3\% of the data is modified via delete and insert operations, resulting in new files being added to the table. 
This significantly degrades performance in the subsequent \emph{single-user} phase, increasing execution time by a factor of 1.53$\times$. 
However, manually triggering compaction restored performance to levels comparable to the initial execution of the workload. 
This experiment highlights that effective data maintenance is not only necessary for the storage layer but also significantly impacts query execution robustness and efficiency.
As a drawback, note that compaction running concurrently with a user's workload may cause write-write conflicts, resulting in longer execution time due to potential retries and wasted resources.
However, we experimentally show in \S\ref{sec:synt_eval} that the benefits usually outweigh the cost.

\smallsection{Limitations of Manual Intervention} 
Although compaction has proven effective, manually selecting tables for compaction is clearly not scalable to meet LinkedIn's operational demands. 
Specifically, to address the small file problem, data infrastructure engineers had been dedicating increasing amounts of time to developing cost-effective strategies for reorganizing onboarded data, while ensuring scalability and managing the maintenance and onboarding of additional tables. 
This reactive approach is unsustainable, as it allows user workflows to fail before compaction can be applied.  
At the same time, enabling periodic compaction across the entire fleet of 21K onboarded tables in OpenHouse (projected to grow to 100K by next year) was also determined to be prohibitively expensive in terms of both capital expenditure (capex) and operational expenditure (opex). 
Our analysis quantified the magnitude of this challenge: compacting approximately 3K raw event tables in the managed ingestion pipeline uses a daily average compute capacity of 150TBhrs, and a daily peak compute capacity of 600TBhrs.
As a result, we started to develop \framework as a \emph{resource-conscious} way of enabling compaction in LinkedIn, initially executing it over a limited selection of tables.
Its effectiveness even within a short timeframe is shown in~\Cref{fig:back:oh-distribution}, allowing OpenHouse to shift the file size distribution towards the target file size at an accelerated pace since its rollout, as discussed further in~\S\ref{sec:impactpractice}.
In contrast, running compaction on a fixed set of tables at a predefined frequency did not yield a significant impact on file size distribution, especially once the system reached a quasi-normal state, leading to fewer opportunities for further optimization.
As a result, subsequent compaction runs often processed files that were already well-sized and balanced, yielding minimal improvements in file size distribution.
This diminished return highlights the inefficiency of static compaction schedules that do not adapt based on the dynamic changes in data patterns and table usage as well as the need for automatic table selection and compaction capabilities.

\section{\framework Overview}
\label{sec:overview}

Building on the conclusions drawn from the previously discussed scenario, our objective is to design and implement a framework that enables automatic data compaction in production environments, carefully balancing its benefits and associated costs. 
Our design is guided by a set of functional (\functional{}) and non-functional (\nonfunctional{}) requirements, which we outline in the following sections. 
We then introduce \framework, our proposed solution.

\subsection{Functional Requirements}
Our functional requirements define the necessary capabilities that a framework must have to effectively address the identified challenges for auto-compaction. 

\vspace{1mm}
\noindent \functional{1}:
\textit{Fine-grained work units.}
\framework should automatically select compaction candidates based on dynamic data analysis.
It should also identify fine-grained work units to execute compaction at the optimal level of granularity, maximizing potential benefits.
By providing the option to break down compaction workloads into smaller, sub-table work units that can be processed independently, the framework can effectively distribute the compaction tasks across segments from different large tables.
This approach enhances parallelism and resource use, allowing the system to prioritize the most impactful segments across the large number of tables.
Smaller work units are also easier to schedule and need fewer resources, which is particularly advantageous in resource-constrained environments. 
It ensures incremental progress, enhances fault tolerance by reducing the need for full table restarts after failures or conflicts, and minimizes disruptions to ongoing operations.
However, we must remain aware of the start-up cost of instantiating more compaction tasks.

\vspace{1mm}
\noindent \functional{2}:
\textit{Support for multiple compaction strategies.}
The framework should support various compaction strategies that can encode the benefits, costs, or a combination of both, depending on the optimization objective. 
For instance, to reduce the load on the storage layer, a benefit-based trigger could greedily prioritize tables with a higher number of small files. 
Additionally, in resource-constrained situations, this trigger could be enhanced with cost-awareness to prioritize operations that yield higher benefits at a lower cost. 
Switching between triggers ensures adaptability to diverse scenarios and maintains balance between performance and resource use.

\vspace{1mm}
\noindent \functional{3}:
\textit{Periodic and post-write execution triggers.} 
The framework should support execution triggered both periodically and immediately after large write operations. 
Periodic execution ensures regular data layout optimization, preventing excessive fragmentation over time and offering predictable cost management. 
Post-write execution enables immediate reorganization, improving performance and curbing file proliferation after significant data ingestion.

\subsection{Non-Functional Requirements}
Next, we introduce non-functional requirements that help us design \framework as those requirements that broaden its applicability beyond the scope of our specific use case.

\vspace{1mm}
\noindent \nonfunctional{1}:
\textit{Extensibility.} 
The framework should be designed with future extensibility in mind, enabling it to integrate additional compaction strategies and adapt to new workloads as needed.
This is important due to diversity of data lake workloads, and the ability to mix and match components (e.g.,~compaction strategies, scheduling policies) ensures the system can evolve without major re-engineering.

\vspace{1mm}
\noindent \nonfunctional{2}:
\textit{Explainability.} 
The framework should produce consistent compaction decisions under identical input conditions (e.g.,~file size distribution, workload characteristics). 
Deterministic decision-making simplifies debugging, testing, benchmarking, and documenting the optimizer's behavior in large-scale production environments, making the system more transparent and manageable.

\vspace{1mm} \noindent \nonfunctional{3}: \textit{Cross-platform compatibility.} 
The framework should be designed to work seamlessly across different \lst and catalog implementations. 
This approach extends its utility beyond LinkedIn's use case to other data lake-centric platforms such as Fabric~\cite{fabric}. 
Such flexibility enables the framework to adapt to a wider range of deployment environments, broadening its impact and applicability.

\begin{figure}[t]
    \centering
    \includegraphics[width=\columnwidth]{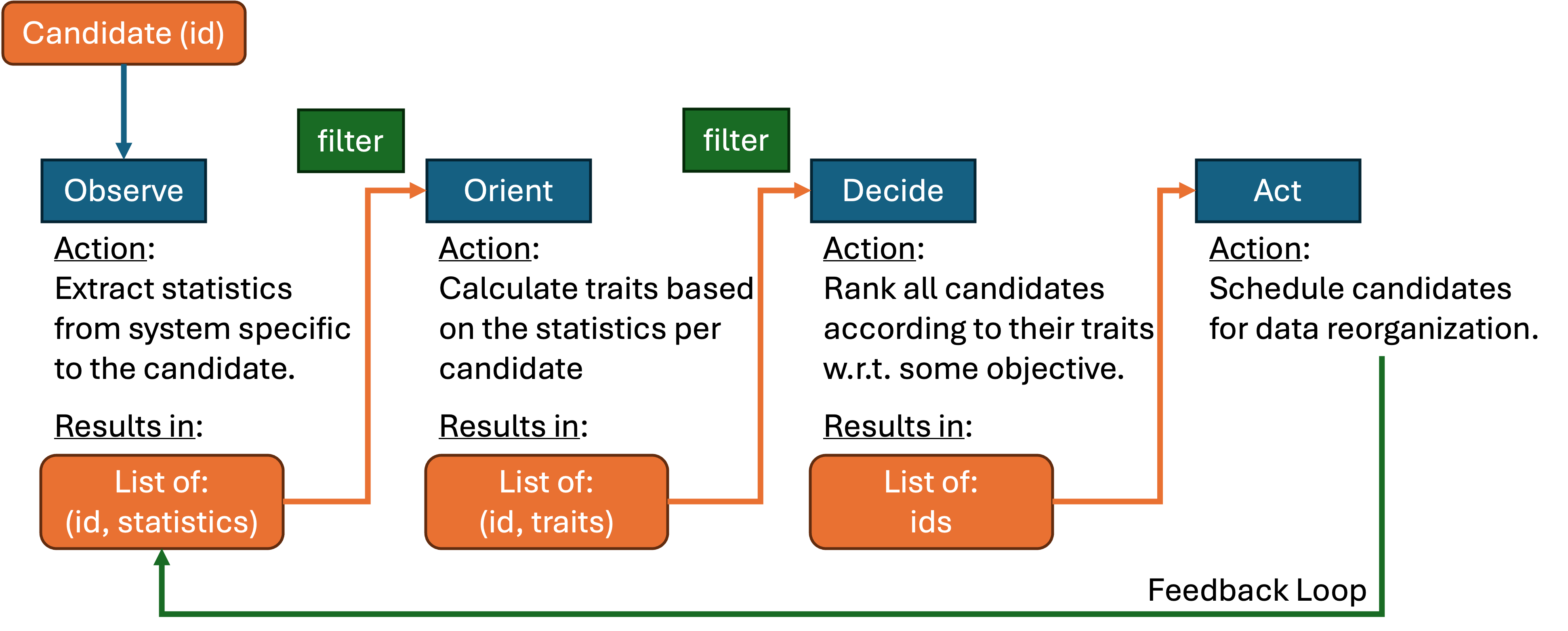}
    \vspace{-1.25em}
    \caption{End-to-end workflow for \framework.}
    \label{fig:architecture:workflow}
    \vspace{-0.5em}
\end{figure}

\subsection{\framework: A Framework for Compaction}
Considering the desiderata previously outlined, we next describe the workflow for a universal, automated compaction framework for \lsts, referred as to \framework in the following. 
We employ a decision-making model known as `Observe, Orient, Decide, Act' (OODA) to map the compaction workflow within the framework. %
which is a model that has been similarly employed in Netflix's auto-optimize functionality~\cite{netflix-auto-optimize}, tailored to their specific use case. 
Our work leverages the same foundational concept but generalizes and modularizes the components of the compaction decision workflow, allowing \framework users to customize it according to their specific requirements. 
As illustrated in \Cref{fig:architecture:workflow}, each of the four phases in the OODA model is associated with an input, an output, and an action that transforms the input into the output.
We first generate compaction candidates that are used in the \textbf{observe} phase to extract relevant statistics needed for downstream decision-making. 
These statistics may include file-level metrics, as well as table or partition-level statistics specific to the candidate. 
A fine-grained approach to candidate generation directly supports \functional{1}, ensuring that compaction tasks can be executed at sub-table levels for efficient resource management.
The output of the observe phase feeds into the \textbf{orient} phase, where the extracted statistics are used to generate \textit{traits}. 
Traits are characteristics that describe either the current state of the candidate or its future potential. 
Examples of traits include file entropy or the estimated computational cost of rewriting data files for compaction. 
The use of traits allows \framework to support \functional{2} by representing different decision strategies, which facilitate the ranking of candidates according to various objectives in the \textbf{decide} phase. 
In this phase, candidates are ranked based on a predefined ranking function, resulting in an ordered list for compaction, which is then processed in the \textbf{act} phase.

Optional filtering mechanisms are optional between the observe and orient phases, as well as between the orient and decide phases, to refine the candidate pool. 
Example filters might check the table size to skip tables that are too small or verify whether a compaction candidate has undergone recent frequent writes to avoid potential conflicts during compaction. 
\framework also supports an optional feedback loop from the act phase back to the observe phase. 
This feedback loop can include updated information such as the new number of partitioned files or layout changes, enabling continuous refinement of the compaction process. 

\framework's architecture supports various modes of operation, including standalone execution on a schedule and proactive use triggered by specific events, aligning with \functional{3}. 
This flexibility enables the framework to adapt to different operational needs without significant reconfiguration. 
Furthermore, the modular design of these phases supports \nonfunctional{1}, allowing new compaction strategies or decision criteria to be integrated seamlessly as long as the data exchanged between phases maintains a consistent structure. 
In addition, by choosing deterministic algorithms for each phase, \framework can address \nonfunctional{2}, making decision-making transparent and predictable. 
Finally, \framework can interface with different catalogs or \lsts through connectors that feed data into the system according to a consistent data model. 
This approach fulfills \nonfunctional{3}, enhancing the framework's reusability and extending its applicability to different data lake-centric platforms. 
Guided by \framework's workflow, we detail crucial implementation details necessary for building our compaction framework in \S\ref{sec:details}. 
We then discuss execution strategies for triggering compaction in \S\ref{sec:auto-datacompact}.

\section{Implementation Details of \framework}
\label{sec:details}

This section covers identifying eligible entities for compaction, referred to as \textbf{candidate generation} (\S\ref{subsec:candidategen}), our approach to \textbf{trait generation}, which outlines how system statistics are utilized effectively (\S\ref{sec:arch:trait}), methods for objective-oriented \textbf{ranking} of candidates (\S\ref{sec:framework:ranking}), and the \textbf{scheduling} of selected candidates (\S\ref{sec:framework:scheduling}). 

\subsection{Generation and Filtering of Candidates}\label{subsec:candidategen}
In the following, we term a \textit{candidate} a collection of files to be compacted. 
While this could represent an entire table, the scope of candidates can be adjusted to fit partitions or snapshots either manually or automatically. 
For example for larger tables, scoping candidates at the partition level enables parallel processing of multiple compaction tasks. 
Adjusting the scope to the snapshot level is particularly beneficial when (reasonably) fresh data needs more frequent access, ensuring performance objectives are met for a subset of the data. 
Candidates can be generated for a single scope or a combination of scopes within the workflow. 
Triggering the workflow for a single scope simplifies the downstream scheduling phase by eliminating the need to manage overlapping scopes. 
However, it is less flexible than considering the entire candidate space, as different table layouts may benefit from different scoping strategies. 

Once candidates are generated, filtering mechanisms are applied throughout the workflow to refine the exhaustively generated candidate pool based on statistics and current table usage. 
The challenge to address is understanding how the tables containing these candidates are being utilized and applying filters accordingly. 
For example, we need to consider the impact of table deletions, table overwrites, or the creation of a table as an `intermediate table' to avoid redundant or conflicting efforts.
These filtering steps are specific to the platform where the framework is deployed and depend on the engines executing the workloads.
For example in OpenHouse, we ensure that tables are not compacted if they have been created recently, i.e.,~within a preset time window.
This approach enables us to avoid spending the computation budget on tables that are not going to affect the long-term health of the system.

Similarly, the specifics of extracting statistics during the observe phase are also dependent on platform characteristics. 
To modularize this step, we propose a standardized layout for statistics that accommodates both generic and custom metrics.
Examples of generic statistics include the number of files in a candidate as well as their corresponding file sizes. 
Custom statistics, on the other hand, could include candidate access patterns and usage metrics--information that may not be available in all systems.

\subsection{Trait Generation}\label{sec:arch:trait}
The second phase, orient, uses statistics collected during the observe phase to calculate so-called \emph{traits} that act as decision helpers for prioritizing and ranking candidates in the next step. 
Traits in \framework are defined independently of one another and can be partially combined during ranking. 
In our work, we primarily focus on two categories of traits: those describing the \textit{benefit} of compaction, such as \textbf{file count reduction} and \textbf{file entropy}~\cite{netflix-auto-optimize}, and those representing its \textit{cost}, such as \textbf{compute cost}. 
This combination enables a cost-benefit analysis to determine the most effective candidates for compaction.

\smallsection{File Count Reduction} 
For a given compaction candidate $c$, we estimate file count reduction after compaction, denoted as $\Delta F_c$, as:
\begin{small}
\[
\Delta F_c = \sum_{i=1}^{\mathit{FileCount}_c} \mathbf{1}\left(\mathit{FileSize}_{i,c} < \mathit{TargetFileSize}_c\right)
\]
\end{small}

The target file size is a configurable parameter that can be chosen based on factors such as the system setup. 
For instance, in HDFS deployments, it is often set to match the HDFS block size. 
The selection can also be influenced by workload characteristics. 
Further discussion on tuning such parameters is provided in \S\ref{subsec:autotune}.

\smallsection{Compute Cost}
Compaction itself incurs costs that need to be considered, especially in a production environment where the benefit/cost ratio is crucial. 
For instance, if two candidates yield different file count reductions (e.g., 200 files versus 100 files) but share similar compute costs, the table with the greater reduction should be prioritized. 
However, if the compute cost for the first candidate is significantly higher--perhaps due to larger file sizes--the benefit/cost ratio may favor the second candidate. 
In resource-constrained scenarios, compaction tasks must be managed within available capacity. 
Candidates with a compute cost that exceeds the allocated budget can be either automatically discarded or flagged for further review if the potential benefit justifies the higher cost. 

To estimate the compute resources required to compact a candidate $c$, denoted as $\mathit{GBHr}_c$, we use:
\begin{small}
$$\mathit{GBHr}_c = \mathit{ExecutorMemoryGB} \times \left(\frac{\mathit{DataSize}_c}{\mathit{RewriteBytesPerHour}}\right)$$
\end{small}

\noindent where $\mathit{ExecutorMemoryGB}$ is the memory allocated to executors for processing the compaction task, $\mathit{DataSize}_c$ is the sum of the candidate files in bytes, and $\mathit{RewriteBytesPerHour}$ indicates the system's throughput in terms of bytes that can be processed per hour.
While this model focuses on executor memory, additional factors--such as compute units, disk, and network I/O--are left for future work.

\subsection{Candidate Ranking and Selection}\label{sec:framework:ranking}
The core objective of the decide phase is to rank compaction candidates and prioritize them for execution. 
We consider two main scenarios for ranking, \textbf{unconstrained} resource availability and \textbf{resource-constrained} compaction systems. 

\smallsection{Unconstrained Resource Scenario} 
When \framework operates without resource constraints, ranking is simplified to a decision function that selects candidates for (immediate) compaction when specific traits exceed predefined thresholds. 
For instance, an engine focused on maintaining optimal query performance might set a target to trigger compaction when the estimated file count reduction, $\Delta F_c$, reaches at least 10\%. 
In this scenario, when a table update occurs, candidates and their traits are generated during the observe and orient phases. 
If $\Delta F_c$ for any candidate indicates a potential file count reduction of 10\% or more, the candidate is passed to the act phase for prompt execution. 
While this approach minimizes file count proactively and enhances user performance, it may also lead to inefficient resource use, particularly for temporary or non-critical tables (though custom filtering rules can be enabled in \framework to mitigate this). 
Additionally, frequent compactions can drive up resource costs, making this approach unsuitable for certain production environments. 

\smallsection{Resource-Constrained Scenario} 
When \framework operates in environments where resources must be carefully managed, we propose ranking candidates based on a combination of traits to balance trade-offs, such as maximizing file count reduction while minimizing compute cost, and to align compaction tasks with available capacity. 
We formalize the candidate ranking process as a Multi-Objective Optimization Problem (\textsc{MOOP}) and scalarize it into a single-objective function using a weighted sum to simplify prioritization. 
To facilitate consistent comparisons, each trait is first normalized using min-max normalization:
\begin{small}
$${T'}_{i,c} = \frac{T_{i,c} - \text{min}(T_i)}{\text{max}(T_i) - min(T_i)}$$
\end{small}

\noindent where $T_{i,c}$ represents the actual value of trait $i$ for candidate $c$, and ${T'}_{i,c}$ is its normalized value.
This normalization scales trait values to a range of $[0, 1]$. 
Next, we define weights $w_i$ for the objectives, ensuring that $\sum(w_i) = 1$. 
These weights indicate the relative importance of each trait within the MOOP function and can be adjusted dynamically to reflect current priorities.
As an example, consider a MOOP function that maximizes the file count reduction while minimizing the associated compute cost as pointed out above.
Here the scalarized score for a candidate $c$ is expressed as:
\begin{small}
$$S_c = w_1 \times T'_{1,c} - w_2 \times T'_{2,c}$$
\end{small}

\noindent where $T'_{1,c}$ represent normalized file count reduction and $T'_{2,c}$ normalized compute cost. 
Candidates are then ranked in descending order based on $S_c$, with higher scores indicating better overall performance relative to the specified objectives. 
To determine the available compute budget based on the cluster's characteristics, \framework can calculate it using available resources such as $\mathit{ExecutorMemoryGB}$ and the predicted time to compact the chosen candidates.
Alternatively, the compaction budget may vary depending on the production environment.
For example, some production systems may instead use a fixed budget determined by capex and organizational limits to ensure to ensure compaction does not exceed preset constraints.
After finalizing the available budget, \framework selects the top-$k$ candidate compaction tasks, where $k$ is the maximum number of candidates that fit within the budget.
Note that the selection function may again differ depending on the production system; however, a reasonable greedy heuristic is to fit as many high-priority compaction tasks as possible within the budget.

By integrating these multi-objective considerations into the ranking phase, \framework ensures that compaction decisions are optimized for both performance and resource efficiency, dynamically adapting to operational constraints and shifting priorities.

\subsection{Compaction Scheduling}\label{sec:framework:scheduling}
The final step in the auto-compaction process is scheduling the selected compaction candidates as part of the act phase. 
Depending on the cluster configuration, compaction can be scheduled on the same cluster or offloaded to a dedicated compaction cluster to minimize the impact on user performance caused by high write operation volumes and resource utilization. 
In practice, \framework allows users to customize the scheduler to suit specific cluster needs. 
For example, when compaction runs on the same cluster as user transactions, compaction tasks can be scheduled sequentially to mitigate resource contention or deferred to off-peak hours if usage patterns are predictable. 
Furthermore, the choice of \lst also influences scheduling decisions. 
For instance, in our experiments with Apache Iceberg~v1.2.0 and OpenHouse, we observed that, counterintuitively, compaction operations executed concurrently could result in conflicts when targeting distinct partitions within a table, leading to failed compaction attempts. 
Thus, any scheduling algorithm must consider the specific characteristics of the chosen \lst, not only in terms of conflict resolution mechanisms but also task failure, recovery, and checkpointing during compaction~\cite{iceberg-compaction-params}, which can impact scheduling decisions.

\section{Automatic Data Compaction}\label{sec:auto-datacompact}
With candidates for compaction identified, we next need to determine when to trigger compaction. 
Instead of relying on manual intervention, we envision automatic scheduling and execution of compaction operations based on factors such as cluster health, compute resources availability, and other relevant metrics. 
Automatic compaction can be implemented in two different ways: 
($i$)~\textit{Optimize-After-Write}, where a candidate's potential for compaction is evaluated each time its files are modified, and 
($ii$)~\textit{Periodic Compaction}, which runs the compaction workflow at regular intervals, such as once per day, to assess and schedule compaction as part of periodic evaluation of the data lake's state.

\smallsection{Optimize-After-Write}
Several existing architectures~\cite{azure-databricks-hook,aws-iceberg-optimize,polaris-transactions} leverage hooks integrated within the engine to enable automatic compaction in response to write modifications, `pushing' the compaction decision onto the engine. 
The same traits described earlier can be used as triggers; if a trait value surpasses a defined threshold, a compaction operation can either be triggered immediately or the optimize-after-write hook can notify the auto-compaction service that changes have occurred and related candidates' traits need recalculation.
The immediate triggering approach ensures the table remains in an optimal state but requires an unlimited compaction budget.
Its alternative, which decouples the hook from scheduling, provides more flexibility in terms of resource usage, allowing for controlled trait generation and efficient compaction task execution.

\smallsection{Periodic Compaction} 
Instead of modifying engine drivers directly, we can choose to implement auto-compaction as a standalone service~\cite{netflix-auto-optimize,hudi-compaction,aws-glue-optimize}, potentially integrated into a catalog or control plane like OpenHouse. 
This service runs independently, periodically evaluating whether compaction criteria are met, `pulling' information about the state of the cluster and scheduling compaction accordingly. 
It is especially advantageous in scenarios with predictable compaction cycles, such as scheduling compaction when cluster utilization is low during off-peak hours or ensuring that compaction does not interfere with other active workloads. 

\begin{figure}[t]
    \centering
    \includegraphics[width=.85\columnwidth]{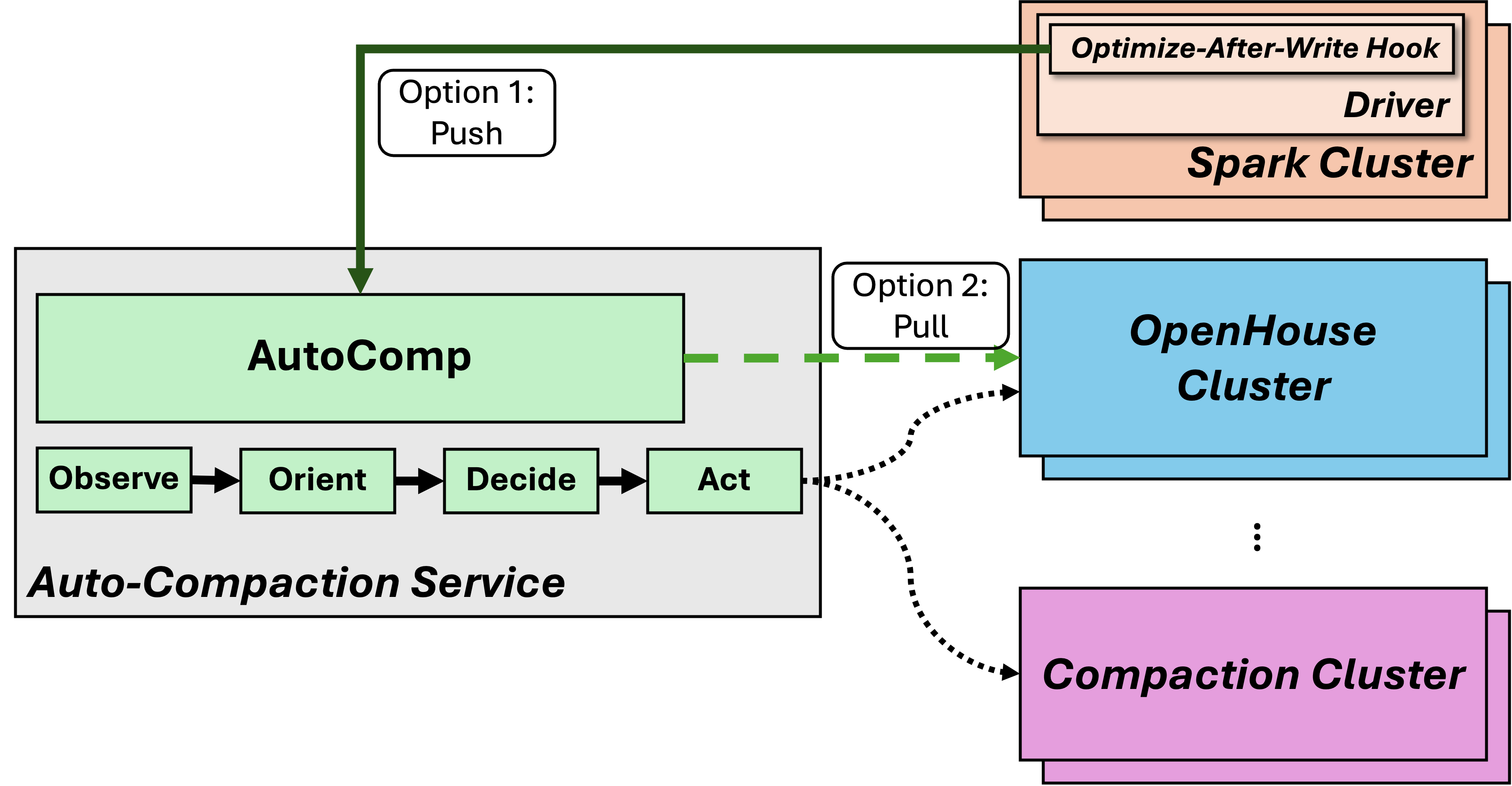}
    \vspace{-0.5em}
    \caption{Cluster integration of \framework.}
    \label{fig:auto-compact:arch}
\vspace{-0.5em}
\end{figure}

\newparagraph
These strategies integrate seamlessly with \framework's workflow, as depicted in \Cref{fig:auto-compact:arch}. 
Here, \framework functions as a standalone component that supports both push and pull operations, allowing for the (re-)calculation of a candidate's traits either triggered by a hook or retrieved periodically from the OpenHouse cluster.

\section{Evaluation of Compaction Framework}
\label{sec:synt_eval}

This section presents the evaluation of \framework using synthetic workloads, focusing on its effectiveness in mitigating the impact of small file proliferation in \lsts. 

\myparagraph{Cluster Infrastructure and Configuration.}
We conducted the experiments on clusters running Apache Spark v3.1.1 and Apache Iceberg v1.2.0 libraries, mirroring the setup used in LinkedIn's production environment. 
We specified standard configurations for both the driver and executor nodes and enabled Adaptive Query Execution (AQE)~\cite{apache-spark-aqe}. 
The query-processing cluster consisted of one driver node and 15 executor nodes, while the compaction cluster used one driver node and three executor nodes. 
Both clusters were provisioned using Azure VMSS, with each node being an Azure Standard E8s v3 instance (Intel\textsuperscript{\textregistered} Xeon\textsuperscript{\textregistered} CPU E5-2673 v4 @ 2.30GHz, 8 virtual cores, 64GB RAM).
OpenHouse v0.5.131 was deployed on a separate Azure Kubernetes Service (AKS) cluster using its default Terraform configuration~\cite{openhouse-terraform}. 
The \framework extension was configured to run periodically and triggered Spark compaction jobs based on its decision logic. 
The data for the experiments was stored in Azure Data Lake Storage Gen2 (ADLS)~\cite{azure-adls}. 
In addition to Iceberg metadata tables~\cite{iceberg-metadata-tables}, we leveraged Logs Analytics~\cite{azure-monitor-log-analytics} to monitor telemetry data across different services.

\myparagraph{Design of Experimental Workloads.}
We used the \textit{CAB-gen} tool~\cite{10.14778/3583140.3583156,cab-gen} to generate metadata for multiple databases and query streams, modeled after real-world usage patterns in cloud data warehouse environments~\cite{snowflake-nsdi20}. 
The database schemas are based on the TPC-H schema, while the query streams mimic usage patterns such as constant demand with sinusoidal variations (e.g., dashboards), short bursts (e.g., interactive queries), large bursts (e.g., daily maintenance jobs), and predictable workloads triggered at specific times (e.g., hourly jobs). 
The \textit{CAB-gen} tool required several parameters: raw \textit{data size}, \textit{number of databases}, \textit{CPU time} (representing total computational workload), and \textit{execution time} (duration of the experiment). 
For our test scenario, we set the parameters to $500$GB of data, $20$~databases, $1$~total CPU hours, and $5$~hours of experiment time. 
After generating the database definitions with \textit{CAB-gen}, we used the \textit{dbgen} tool from the TPC-H benchmark~\cite{tpcds-refresh} to generate synthetic data.
The \textsc{lineitem} table was partitioned by \textsc{shipdate} with monthly granularity, producing a workload with mixed data update patterns across partitioned (\textsc{lineitem}) and non-partitioned (\textsc{orders}) tables\footnote{The original CAB-gen only generated updates on the \textsc{orders} table; we extended this to include updates on both \textsc{orders} and \textsc{lineitem} tables.}.
For query execution, we extended the \textit{LST-Bench} benchmarking tool~\cite{lstbench} to run the streams produced by \textit{CAB-gen}. 
These extensions are now available in OSS LST-Bench~\cite{lstbench-cab}.

\myparagraph{Candidate Selection and Scheduling.}
Our synthetic experiments focus on three different candidate selection strategies:
($i$) no compaction, ($ii$) \textit{table}-scope compaction, and ($iii$) a \textit{hybrid} compaction strategy that chooses partition-scope compaction if the table is partitioned and otherwise defaults to table-scope.
The table-scope compaction mimics the current OpenHouse implementation while the hybrid strategy explores whether partition-based compaction can help to balance the resource utilization load.
Candidates are compacted in parallel on the table level but sequentially on the partition level as we have noticed compaction operations getting dropped due to conflicts even for distinct partitions otherwise, see \S\ref{sec:framework:scheduling} for details.
Compaction execution is triggered every hour of the experiment, i.e.,~a successful experiment should contain four compaction executions in a $5$~hour timeframe.

\myparagraph{Metrics.}
We capture both client- and server-side statistics for a comprehensive understanding of the impact of compaction on workload execution. 
On the client side, we focus primarily on workload query execution times and the number of errors observed during execution.
On the server side, we gather several compaction-related metrics, including current file counts for tables, rewritten bytes, and added files. 
In addition, we compute a custom metric, $\mathit{GBHr}_\mathit{App}$, which reflects the compute resources needed by an application $\mathit{App}$; here, an application is defined at the job level, meaning each triggered compaction operation is treated as a distinct instance.

\begin{figure}[t]
    \centering
    \includegraphics[width=.95\columnwidth]{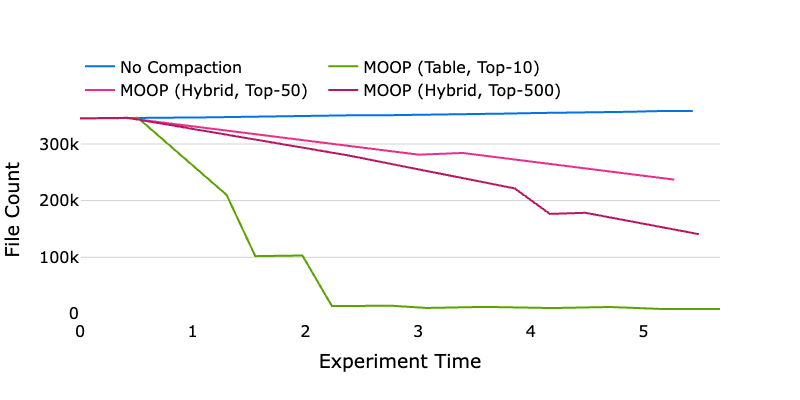}
    \vspace{-1em}
    \caption{Compaction strategy impact on file count over time.}
    \label{fig:exp:file_count_red}
\vspace{-1em}
\end{figure}

\subsection{File Count} 
Our first goal is to evaluate \framework's effectiveness in handling HDFS' \textit{small files} problem in the storage layer.
We run the CAB workload on our query-processing cluster, executing streams for the $20$~databases concurrently while our compaction strategies operated on the separate compaction cluster, triggered at $1$-hour intervals. 
Figure~\ref{fig:exp:file_count_red} shows the file count over time for the baseline with no compaction and for \framework using the \textsc{MOOP} strategy that balances the benefit of reducing the estimated number of small files against the cost of rewriting a table or partition.
We set $k$--the number of work units compacted in each \framework run--to $10$ for table-scope compaction and $50$ resp. $500$ for the hybrid compaction strategy, the target file size to $512$~MB, and the weights for MOOP to 0.7 (file count reduction) and 0.3 (computation cost), mimicking our OpenHouse deployment.
Note that in practice, we may choose to vary the value of $k$ depending on constraints such as available compaction resources or for a gradual rollout in a production environment.
The values chosen for visualization here exemplify trends that we can see across a range of $k$ values.

\myparagraph{Storage Layer Changes.} 
In our baseline (no compaction), we observe a high initial file count, as the data load operation generates many small files--a common scenario in practice due to factors like cluster misconfiguration (\S\ref{sec:motivation}). 
During the experiment, the file count increases steadily, with an average increase of approximately 2,640 files per hour, although the exact number fluctuates with the write queries executed during each interval.
On average, the experiment runs for about five hours, with a noticeable spike in data write operations around hour four due to workload patterns that increase load on the query-processing cluster during that time window. 
With compaction enabled, we observe a significant reduction in file count across all compaction strategies, with an initial sharp decline in file count followed by a more gradual flattening of the curve.
For the \emph{hybrid} strategies, the reduction curve is less steep, as fewer entities are compacted in each round, leading to more gradual, controlled reduction in file count.

\begin{figure}[t]
    \centering
    \includegraphics[width=.95\columnwidth]{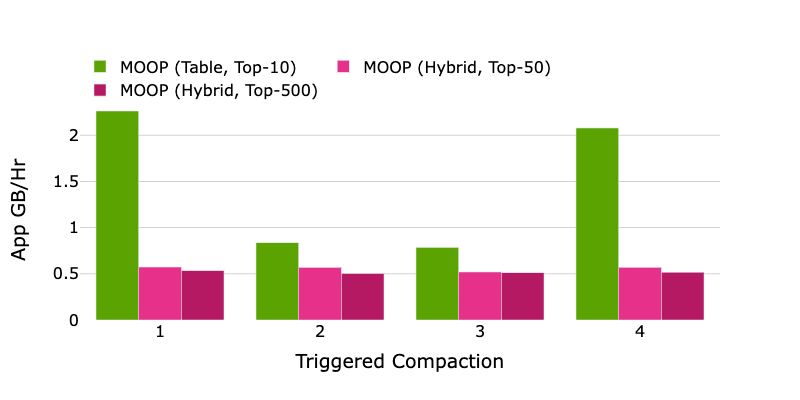}
    \vspace{-1em}
    \caption{Mean $\mathit{GBHr}_\mathit{App}$ for various compaction strategies.}
    \label{fig:exp:meanappgbhr}
\end{figure}

\myparagraph{Compaction Cost.} 
While each compaction strategy reduces file count, it is also important to consider the cost associated with compaction. 
\Cref{fig:exp:meanappgbhr} shows the average $\mathit{GBHr}_\mathit{App}$ for compaction across strategies during the experiment.
Compaction at the \emph{table} level can be advantageous and more effective when a table layout is highly fragmented, but a finer-grained approach, such as the \emph{hybrid} method with partition-level compaction, provides more control, allowing file count reductions at a slower pace and thus balancing resource usage for compaction over time as documented with a more stable value for $\mathit{GBHr}_\mathit{App}$ across compaction operations.

\begin{figure}[!tp]
\setlength{\belowcaptionskip}{0pt}
\begin{subfigure}{0.50\columnwidth}
  \centering
  \includegraphics[width=\columnwidth]{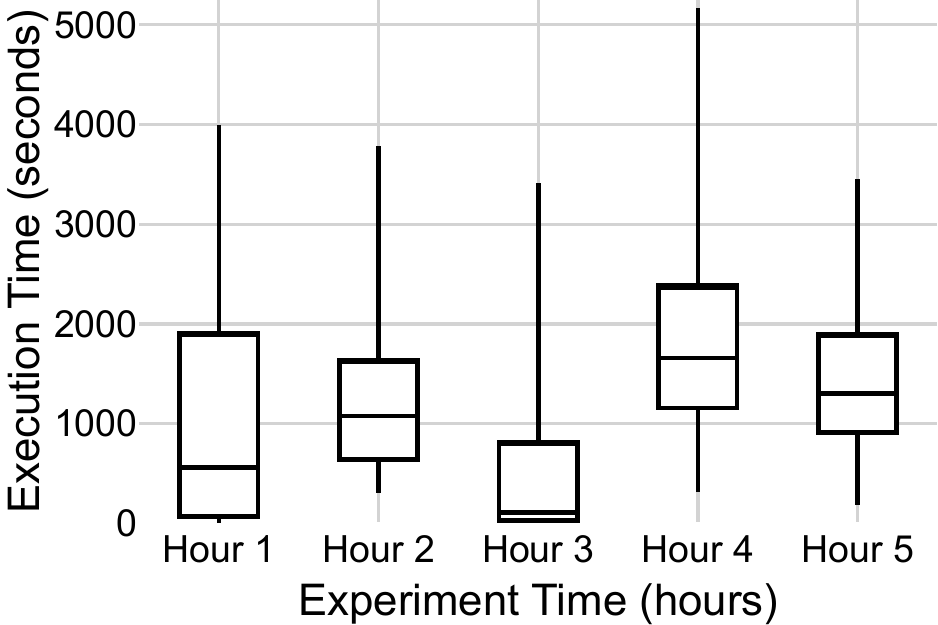}  
  \caption{\footnotesize No Compaction - RO}
  \label{fig:nocompaction-readonly}
\end{subfigure}
\hfill
\begin{subfigure}{0.49\columnwidth}
  \centering
  \includegraphics[width=\columnwidth]{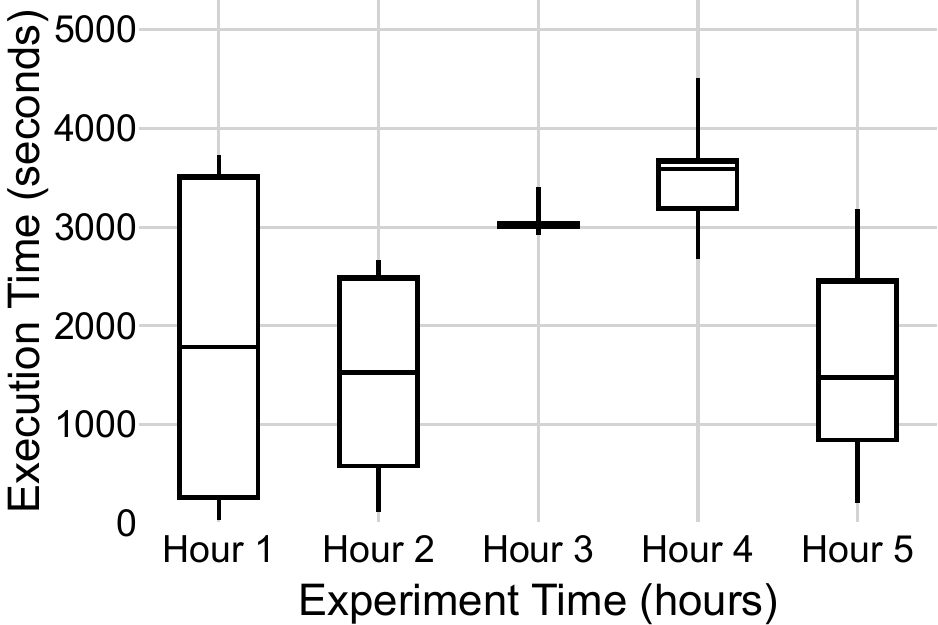}  
  \caption{\footnotesize No Compaction - RW}
  \label{fig:nocompaction-readwrite}
\end{subfigure}

\begin{subfigure}{0.49\columnwidth}
  \centering
  \includegraphics[width=\columnwidth]{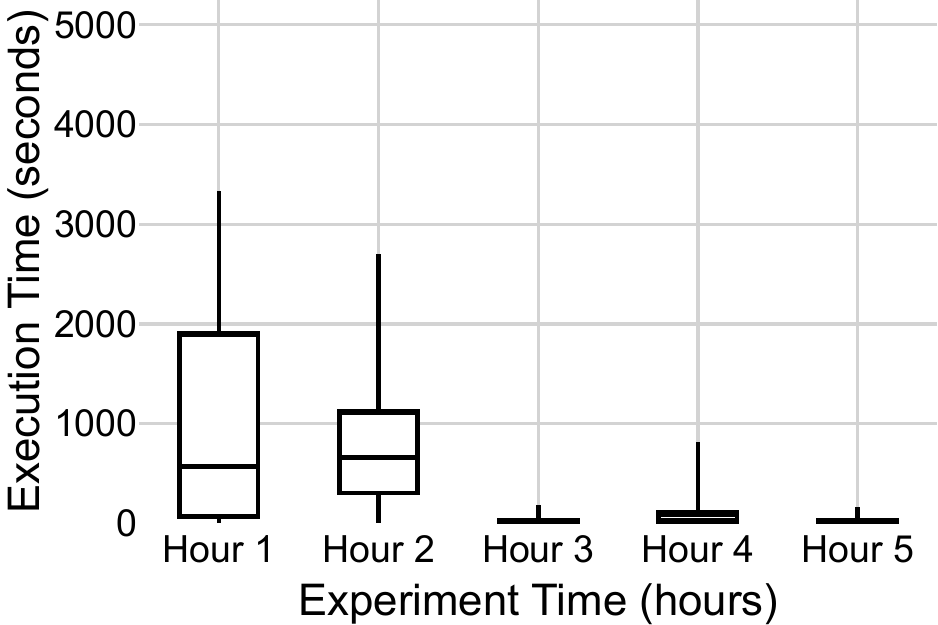}  
  \caption{\footnotesize MOOP (Table, Top-10) - RO}
  \label{fig:hybrid100-readonly}
\end{subfigure}
\hfill
\begin{subfigure}{0.49\columnwidth}
  \centering
  \includegraphics[width=\columnwidth]{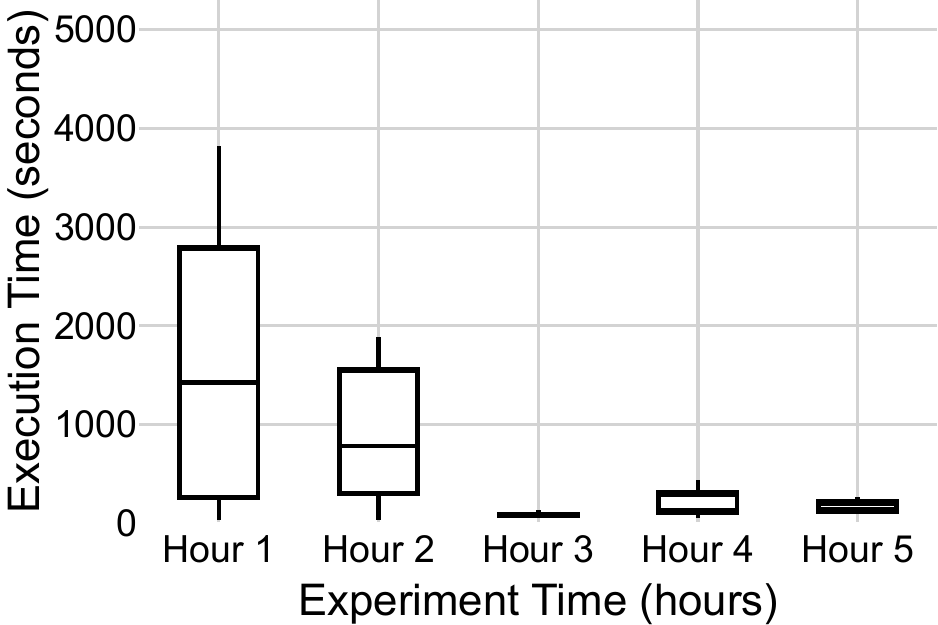}  
  \caption{\footnotesize MOOP (Table, Top-10) - RW}
  \label{fig:hybrid100-readwrite}
\end{subfigure}

\begin{subfigure}{0.49\columnwidth}
  \centering
  \includegraphics[width=\columnwidth]{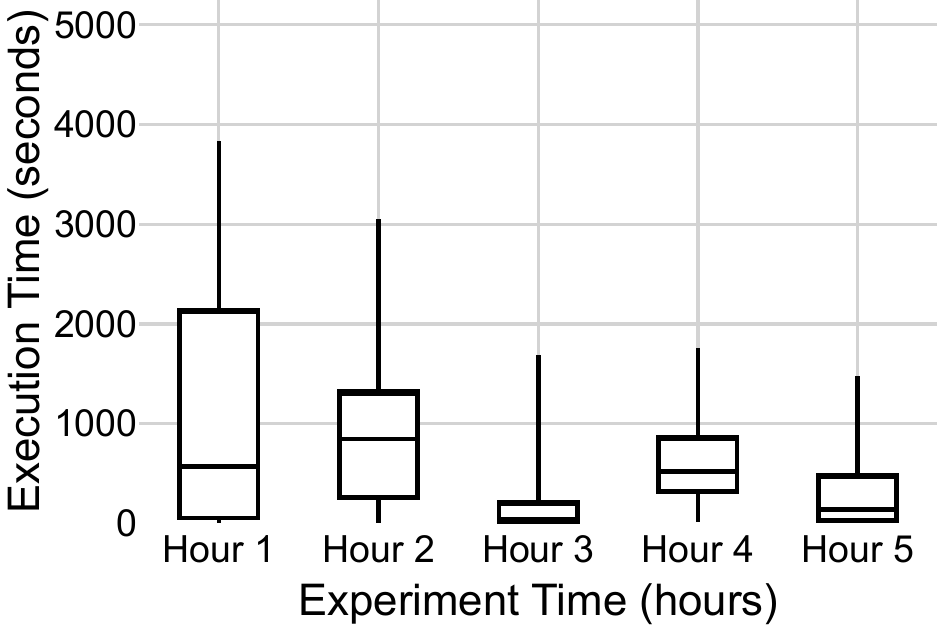}  
  \caption{\footnotesize MOOP (Hybrid, Top-500) - RO}
  \label{fig:hybrid500-readonly}
\end{subfigure}
\hfill
\begin{subfigure}{0.49\columnwidth}
  \centering
  \includegraphics[width=\columnwidth]{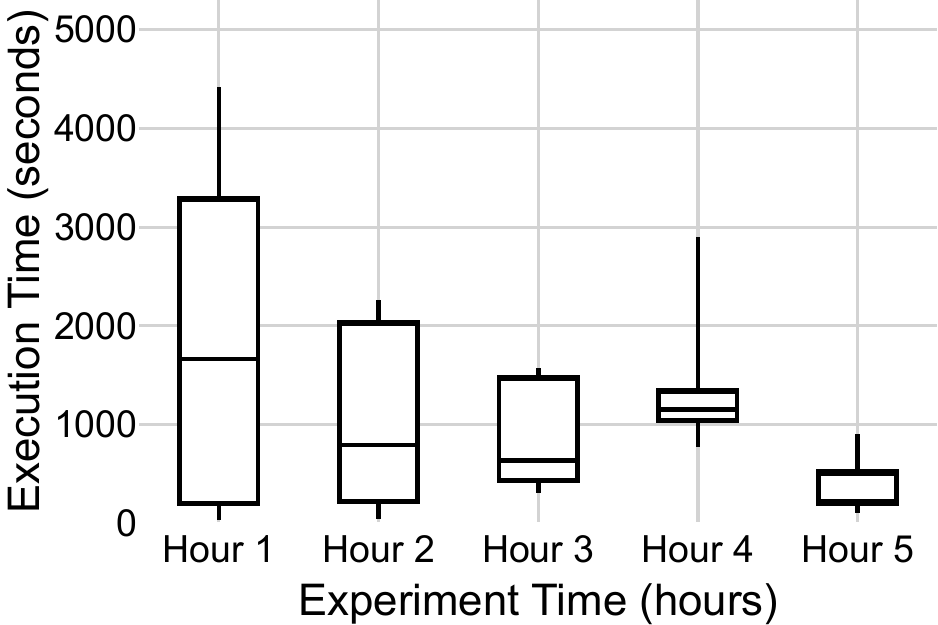}  
  \caption{\footnotesize MOOP (Hybrid, Top-500) - RW}
  \label{fig:hybrid500-readwrite}
\end{subfigure}
\vspace{-0.75em}
\caption{Impact of compaction on query latency.}
\label{fig:exp:query_perf}
\vspace{-0.5em}
\end{figure}

\subsection{Query Performance} 
The proliferation of small files can significantly impact query performance. 
To evaluate \framework's effect on query performance, we measured query execution times over the course of our experiments. 
\Cref{fig:exp:query_perf} shows execution times for both read-only (left column) and read-write (right column) queries under no compaction and various compaction strategies.
Each candlestick bar represents the min, 25th percentile, median, 75th percentile, and max execution times per hour. 
Focusing first on read-only queries, we observe that performance across all strategies is similar in the first hour, but from the second hour onward, compaction consistently improves query performance, with faster reductions in execution times under the more aggressive compaction strategy (\emph{table}, top-$10$). 
Additionally, execution time variability decreases, as shorter query runtimes help reduce resource contention in the query-processing cluster.
For the less aggressive strategy (\emph{hybrid}, top-$500$), we also see significant performance improvements, but not at the level of the more aggressive strategy which is expected and correlates with our previous findings relating to the number of files successfully compacted and thus decreasing the system's file count.
However, we note that the experiment's end-to-end runtime for both compaction strategies meets the pre-set 5-hour limit, while the \emph{no compaction} baseline incurs an additional 25 minutes of overhead due to queuing and longer query execution times.

\begin{table}[t]
\centering
\caption{Client and cluster-side conflicts per execution hour.}\label{fig:exp:retried_queries}
\vspace{-0.5em}
\resizebox{\columnwidth}{!}{
    \begin{tabular}{c|c|c|c|c|c|c}
    \toprule
     \multirow{2}{*}{\textbf{Hour}} & \textbf{\# Write} & \multicolumn{3}{c|}{\textbf{Client-side Conflict}} & \multicolumn{2}{c}{\textbf{Cluster-side Conflict}} \\
     \cline{3-7}
     & \textbf{Queries} & \textsc{NoComp} & \textsc{Table-10} & \textsc{Hybrid-500} & \textsc{Table-10} & \textsc{Hybrid-500} \\
    \hline
    2 & 12 & 1 & \textbf{\textcolor{red}{11}} & 4 & \textbf{\textcolor{red}{23}} & \textbf{0} \\
    3 & 5 & \textbf{0} & 2 & \textbf{0} & \textbf{\textcolor{red}{17}} & \textbf{0} \\
    4 & 15 & 1 & 5 & 6 & 4 & \textbf{0} \\
    5 & 8 & 4 & \textbf{0} & 2 & \textbf{0} & \textbf{0} \\
    \bottomrule
    \end{tabular}
}
\vspace{-0.75em}
\end{table}

In addition to the execution time of the workload queries, we also examined their retry behavior due to write-write conflicts as shown in~\Cref{fig:exp:retried_queries}.
Here, we show queries that have been retried due to client-side errors, i.e.,~versioning conflicts that cause a client-side operation to terminate, and cluster-side conflicts that occur during compaction operations.
We observe that conflicts are present even without compaction due to concurrent writes to the same tables and commonly correlate with spikes in workload patterns as shown by the number of write queries issued within a specified hour.
An exception is our experiment for table-scope compaction.
We observe that conflicts occur early during the experiment due to a large number of compaction operations and subsequent conflicts about stale metadata.
By the fifth hour, the tables with write activity--namely, \textsc{lineitem} and \textsc{orders}--have been largely compacted, leading to a reduction in write-write conflicts.
Interestingly, we observe no cluster-side conflicts for our hybrid approach, suggesting that the probability of disrupted compaction operations decreases with the size of the candidate to be compacted which is expected.

\subsection{Auto-Tuning Compaction Triggers}\label{subsec:autotune}

\begin{figure}[!tp]
\setlength{\belowcaptionskip}{0pt}
\begin{subfigure}{0.49\columnwidth}
  \centering
  \includegraphics[width=\columnwidth]{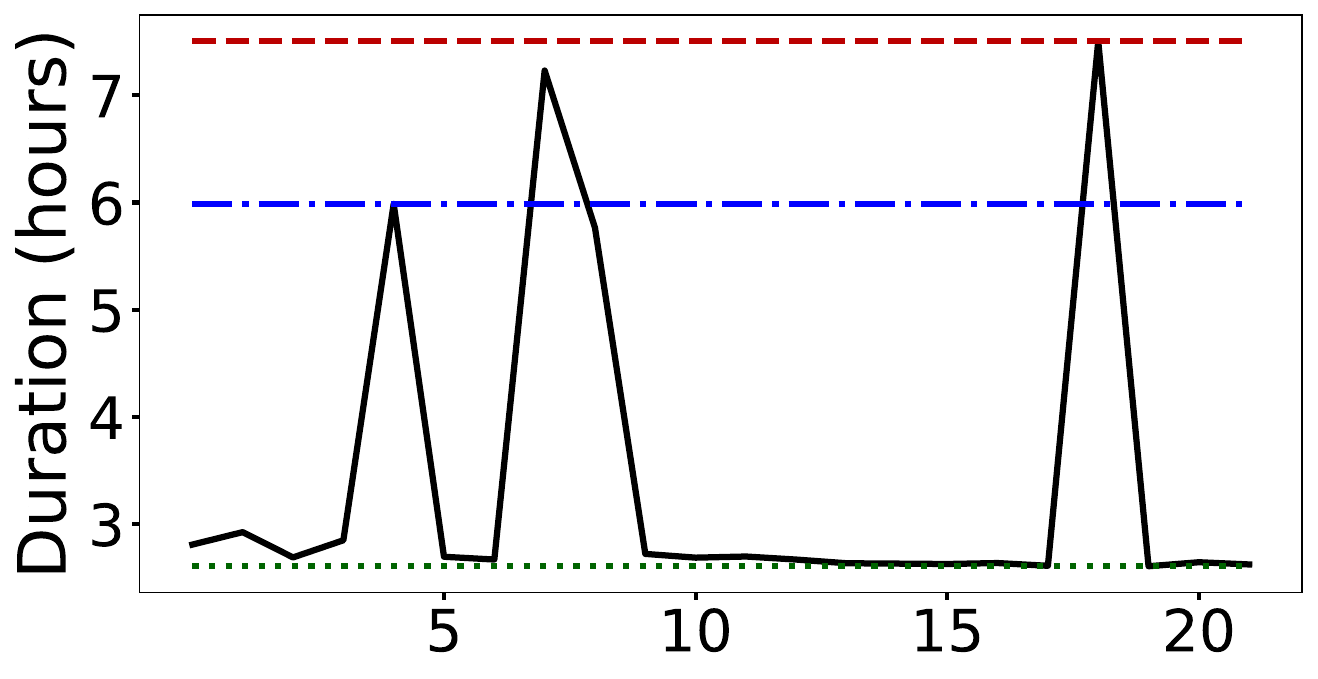}  
  \captionsetup{aboveskip=0pt}
  \captionsetup{belowskip=3pt}
  \caption{TPC-DS WP1, File Count}
  \label{fig:auto-datacompact:wp1-file}
\end{subfigure}
\hfill
\begin{subfigure}{0.49\columnwidth}
  \centering
  \includegraphics[height=2.13cm]{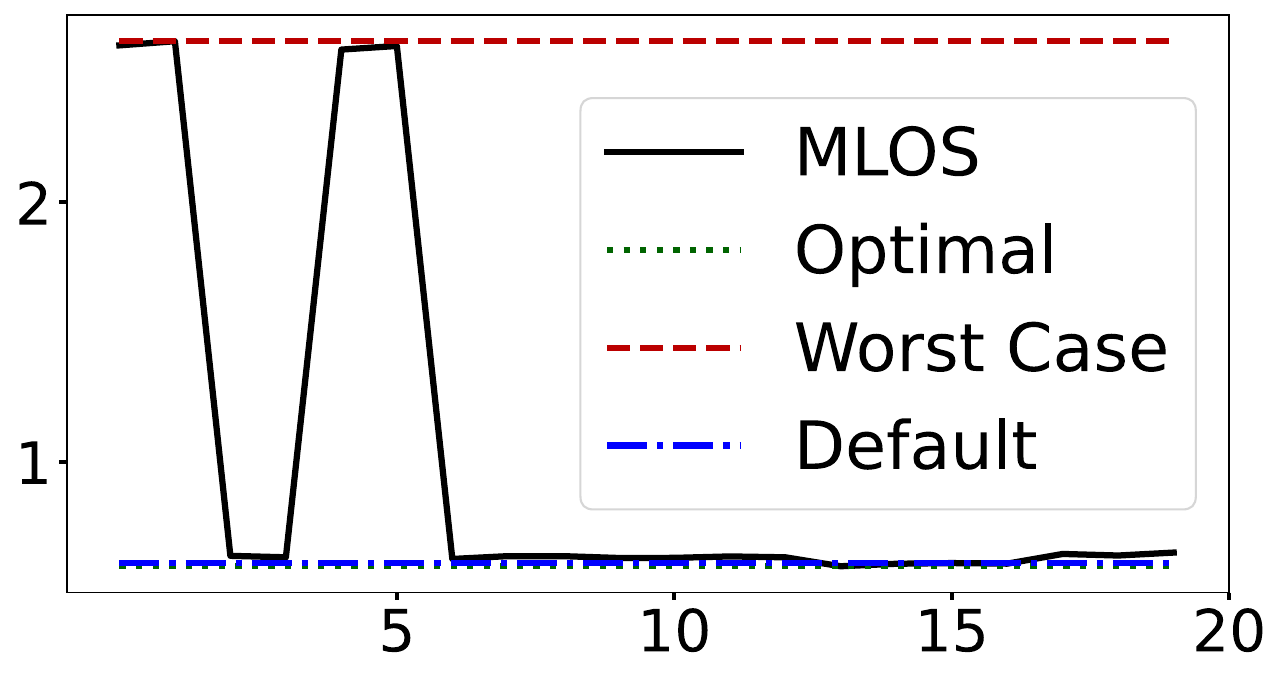}  
  \captionsetup{aboveskip=0pt}
  \captionsetup{belowskip=3pt}
  \caption{TPC-H, File Count}
  \label{fig:auto-datacompact:tpch-file}
\end{subfigure}

\begin{subfigure}{0.49\columnwidth}
  \centering
  \includegraphics[width=\columnwidth]{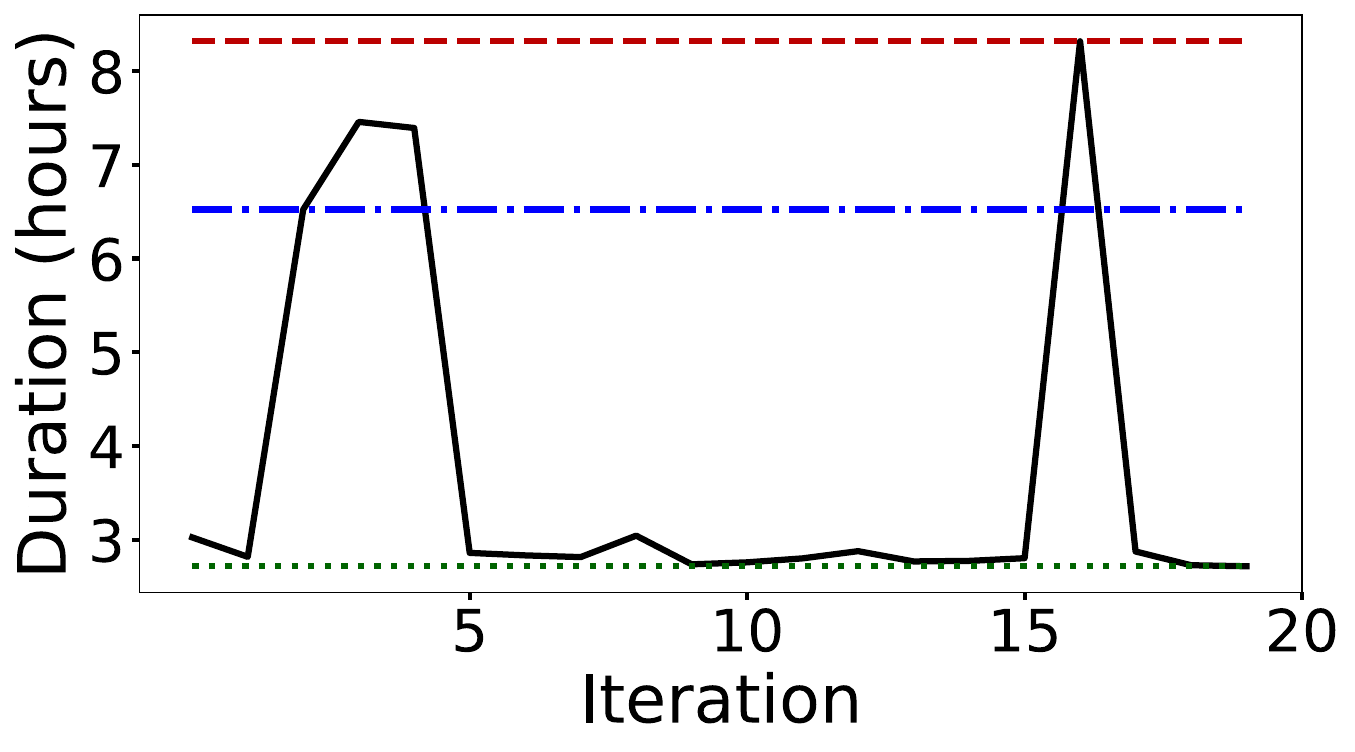}  
  \captionsetup{aboveskip=1pt}
  \captionsetup{belowskip=2pt}
  \caption{TPC-DS WP1, Entropy}
  \label{fig:auto-datacompact:wp1-entropy}
\end{subfigure}
\hfill
\begin{subfigure}{0.49\columnwidth}
  \centering
  \includegraphics[width=\columnwidth]{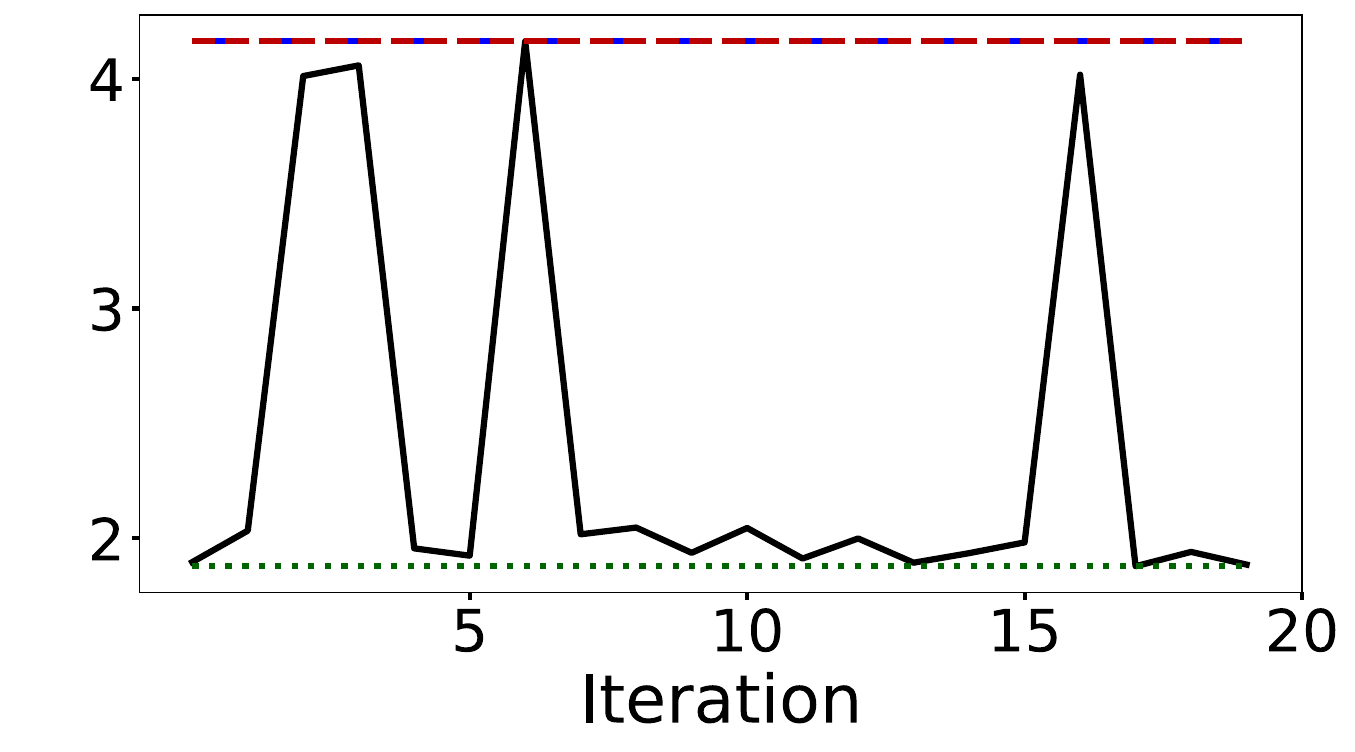}  
  \captionsetup{aboveskip=1pt}
  \captionsetup{belowskip=2pt}
  \caption{TPC-DS WP3, File Count}
  \label{fig:auto-datacompact:wp3-file}
\end{subfigure}
\vspace{-1.0em}
\caption{Comparison of compaction decisions and results.}
\label{fig:auto-datacompact:comparison}
\vspace{-0.5em}
\end{figure}

As noted earlier, a key challenge in auto-compaction is determining parameter values that best fit a given workload. 
Thus, we experiment with an auto-tuning framework in conjunction with \framework, using a simplified optimize-after-write hook setup, i.e.,~unlimited compaction resources. 
We use two compaction traits--small file count and file entropy~\cite{netflix-auto-optimize}--and tune the thresholds that determine when compaction is triggered. 
As before, we deploy LST-Bench with three of its in-built workloads: TPC-DS WP1, a long-running workload with frequent data modifications; TPC-DS WP3, where one compute cluster handles all writes while another handles all reads; and TPC-H. 
Both TPC-DS and TPC-H datasets use a scale factor of 100, with experiments running on a 16-node Spark cluster (plus a 7-node sidecar cluster for writes in TPC-DS WP3), using Delta Lake v2.4.0 as the \lst. 
This demonstrates \framework's flexible design, enabling support for various \lst implementations.
To optimize parameters, we leverage the FLAML optimizer~\cite{wang2021flaml} implemented within MLOS~\cite{mlos,DBLP:journals/pvldb/KrothMAZGT24}, an open-source optimization framework, to iteratively refine threshold values.
\Cref{fig:auto-datacompact:comparison} presents the results, with the y-axis showing total end-to-end experiment duration and the x-axis representing iterations, each with a threshold selected by MLOS, leading to the following observations. 
($i$) For TPC-H (\Cref{fig:auto-datacompact:tpch-file}), the default setting (no auto-compaction) performs best, as compaction rewrites entire non-partitioned tables, making it costly, and its long data modification phase already dominates execution time. 
In contrast, TPC-DS WP1 (\Cref{fig:auto-datacompact:wp1-file}) benefits from compaction when tables become too fragmented, reducing query time by up to 2$\times$ when applied appropriately.
Finally, TPC-DS WP3 (\Cref{fig:auto-datacompact:wp3-file}) sees consistent benefits from compaction, as its decoupled read and write clusters minimize resource contention with other queries. 
($ii$) We observe similar query performance when using small file count- and entropy-based triggers, as shown in \Cref{fig:auto-datacompact:wp1-file} and \Cref{fig:auto-datacompact:wp1-entropy}. 
This suggests both decision functions can yield comparable results depending on the chosen thresholds. 
Note that here, we experiment with single-trait decision functions only; more complex approaches, such as multi-objective ranking functions that consider computation cost, may lead to different outcomes.

\newparagraph
Overall, these experiments demonstrate that auto-tuning a compaction framework like \framework is a promising direction for future exploration, which we discuss further in \S\ref{sec:future}.

\section{\framework Impact in Practice}
\label{sec:impactpractice}

As discussed in~\S\ref{sec:motivation}, the initial solution to LinkedIn's small file problem was manual intervention via compaction tasks.
However, this approach quickly proved infeasible at scale. 
To address the challenge of compacting over \( 35K \) tables in LinkedIn's OpenHouse deployment, we implemented an instantiation of \framework with the following characteristics. 
First, we scoped compaction at the table level, consistent with the existing manual compaction strategy.
Second, we combined the file count reduction estimator and compute cost calculator introduced in~\S\ref{sec:arch:trait} within the \textsc{MOOP} ranking function described in~\S\ref{sec:framework:ranking}.
We tuned the \textsc{MOOP} weights to reflect our specific objectives, adjusting the file count reduction weight ($w_1$) based on a database's quota utilization--measured by its total number of files or namespace objects.
Each database represents a logical group of tables associated with a specific tenant:

\begin{small}
$$w_1 = 0.5 \times \left(1 + \left(\frac{\mathit{UsedQuota}}{\mathit{TotalQuota}}\right)\right)$$
\end{small}

\noindent Here, $\mathit{TotalQuota}$ is the HDFS namespace quota (in number of filesystem objects) allocated to a database, and $\textit{UsedQuota}$ is the currently utilized portion. 
Finally, we implemented a periodic scheduling strategy that triggers once daily, selecting a set of $k$ compaction candidates.
Fixing the number of candidates was critical during initial rollout to ensure predictable behavior, a key requirement when introducing a new (automated) mechanism into production.
Over time, we transitioned to dynamically selecting $k$ based on available compaction resources.

Given our experience with both manual and automatic compaction, we share several observations from production deployments: ($i$)~whether manual compaction alone is sufficient in production, ($ii$)~how compaction has shifted file distributions in OpenHouse, and ($iii$)~how auto-compaction impacts user workload execution and HDFS metadata operations.

\smallsection{Diminishing Returns of Manual Compaction}
Our initial mitigation approach was an ad-hoc manual compaction strategy that repeatedly compacted a fixed set of \(k \approx 100\) tables at a high frequency (e.g.,~daily).
These tables were chosen because of their susceptibility to high fragmentation, and early results showed marked improvements: reduced small file counts, improved query performance, and lower storage overhead.
However, these benefits tapered off over time. 
Once small files were merged, further compaction yielded limited gains.
\Cref{fig:back:oh-distribution} shows that the file size distribution remained largely unchanged between the second and third month of manual compaction.
In general, we observe that not all tables benefit equally from compaction in LinkedIn's OpenHouse deployment. 
While some tables significantly benefit from compaction at a low cost, others incur a high cost with minimal benefits.
In practice, identifying high-impact candidates is non-trivial, as users interact with the system on a daily basis by modifying their data, creating new tables, and adjusting workflows.
As a result, manually defining compaction targets is suboptimal, and motivated our shift towards an automated solution.

\begin{figure*}[!tp]
  \centering
  \begin{subfigure}{0.33\textwidth}
    \centering
    \includegraphics[width=\textwidth]{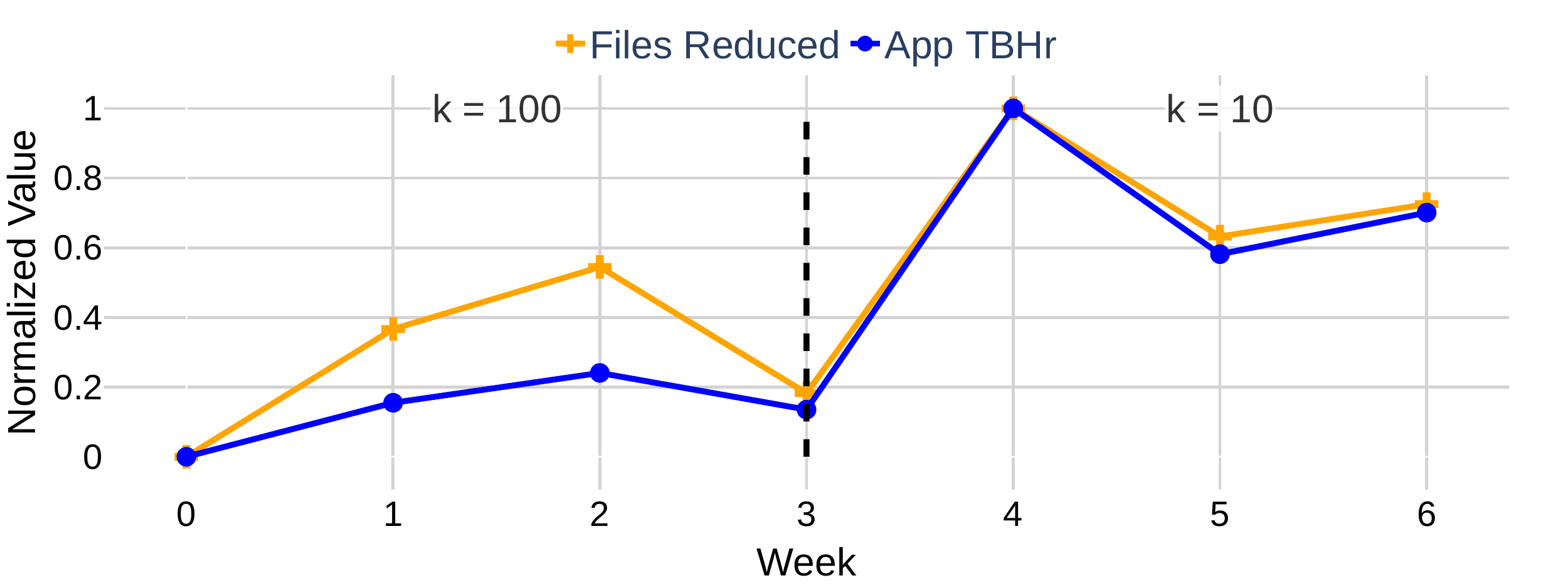}
    \vspace{-1em}
    \caption{Files reduced and computation cost.}
    \label{fig:pract:overall:files_reduced_vs_compaction_cost}
  \end{subfigure}
  \hfill
  \begin{subfigure}{0.33\textwidth}
    \centering
    \includegraphics[width=\textwidth]{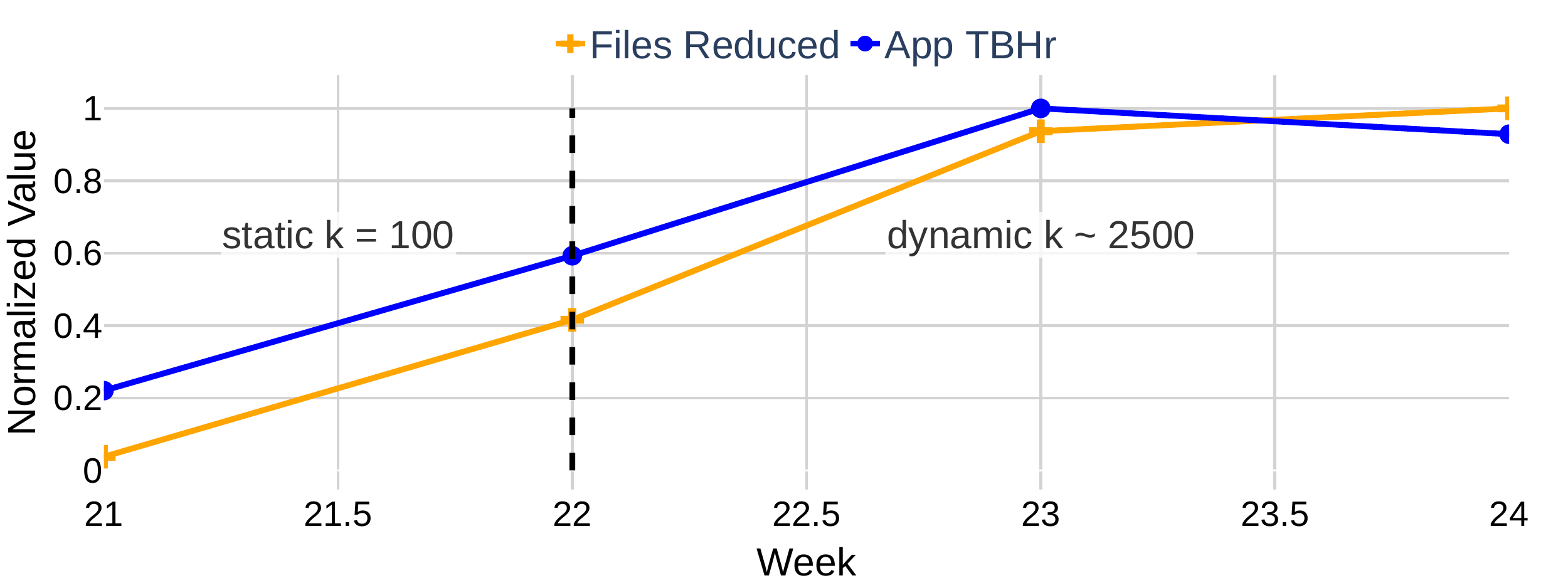}
    \vspace{-1em}
    \caption{Impact of dynamic $k$ tuning.}
    \label{fig:pract:overall:dynamic_k}
  \end{subfigure}
  \hfill
  \begin{subfigure}{0.33\textwidth}
    \centering
    \includegraphics[width=\textwidth]{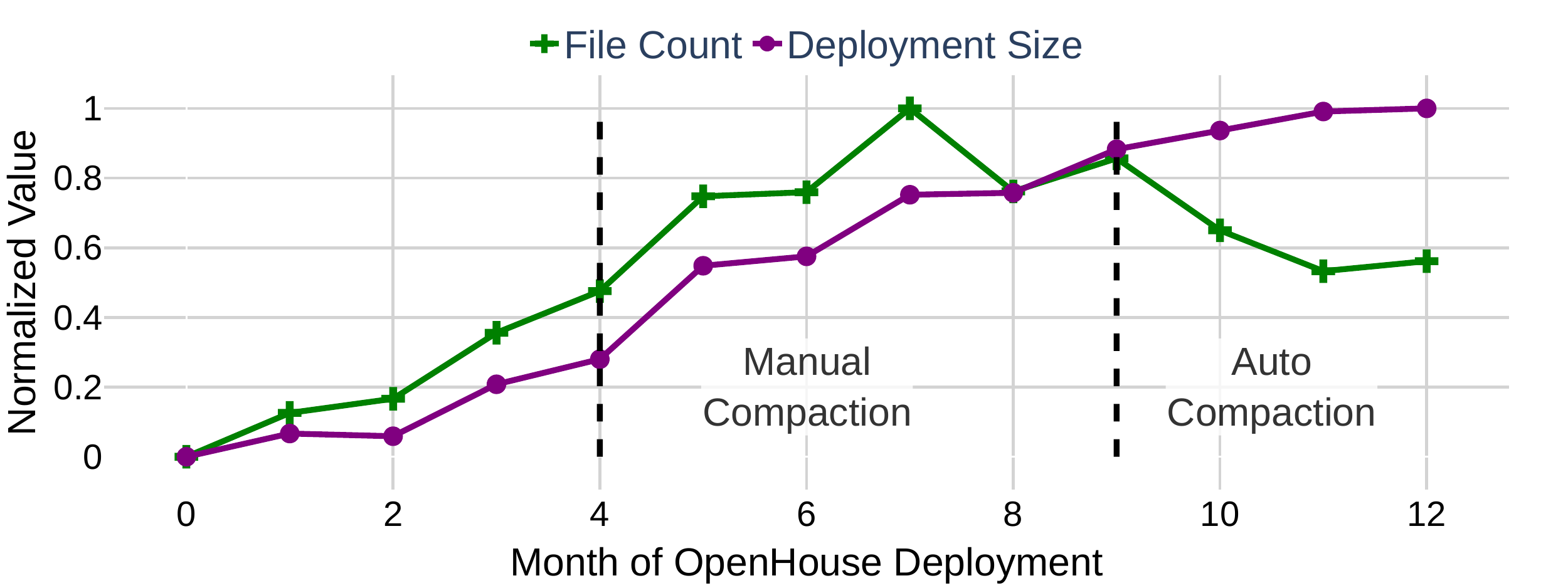}
    \vspace{-1em}
    \caption{Deployment statistics.}
    \label{fig:pract:overall:deployment}
  \end{subfigure}
  \caption{\framework behavior and impact on file count.}
  \label{fig:pract:overall}
\end{figure*}

\smallsection{Deploying Compaction Mechanisms} 
Deploying automatic compaction has significantly alleviated the small file problem in OpenHouse.
Specifically, prior to deployment, users encountered frequent issues, including:
($i$)~query failures caused by HDFS read timeouts due to excessive RPC traffic, 
($ii$)~frequent breaches of user HDFS namespace quotas, and
($iii$)~rapid growth in object count, requiring frequent HDFS federations to distribute the load.
As shown in our motivating example,~\Cref{fig:back:oh-distribution}, prior to compaction tasks being executed regularly, 83\% of the system's files were smaller than 128MB. 
When we introduced manual compaction, we saw a significant shift in overall file distribution with the percentage of small files dropping from 83\% to 62\%.
We further reduced this number by gradually rolling out \framework as part of the OpenHouse compaction decision mechanism.
We began by deploying \framework with a highly conservative choice of $k$, i.e., \(k \approx 10 \), to closely examine the impact of compaction in our production environment without disrupting users.
Interestingly, we observed that switching from manual top-$100$ compaction to automatic top-$10$ compaction strategy effectively increased overall file count reduction, even though we compacted $10\times$ fewer tables.
More specifically, we observed an average reduction of 6.59 million files via manual  compaction versus 7.44 million using \framework with a top-10 selection--an improvement of 12\%.
\Cref{fig:pract:overall:files_reduced_vs_compaction_cost} further shows the relationship between file count reduction and compaction cost (measured in App TBHr) over a 6-week period.
The transition from manual ($k$=100) to auto-compaction ($k$=10) was done in week 3, resulting in both higher effectiveness and higher computation cost.
\Cref{fig:pract:overall:dynamic_k} illustrates the transition in week 22 of the auto-compaction deployment from fixed to dynamic \( k\) selection, constrained by the maximum allocated compaction budget.
With a budget of \( 226\) TBHr, we successfully compacted around \( k \approx 2500 \) tables per iteration of auto-compaction.
Overall, we observe that despite the growing deployment, auto-compaction mechanisms have significantly decreased the file count in HDFS over time, as shown in~\Cref{fig:pract:overall:deployment}.

\smallsection{Model Accuracy and Estimation Errors}
We evaluated the accuracy of our estimators by comparing predicted and actual values for file count reduction and compute cost. 
Unfortunately, we occasionally observe a discrepancy between these values.
For example, we estimated a compute cost of 108 TBHr for one compaction task, but actually consumed 129 TBHr (a 19\% underestimation), while the file count reduction was overestimated by 28\%.
These mismatches suggest that while the current model is generally effective for ranking, it requires further refinement to improve accuracy--particularly in accounting for partition boundaries, as table-level estimates may overestimate the number of small files that can be merged, since compaction does not cross partitions.

\begin{figure}[!tp]
  \centering
  \begin{subfigure}{0.95\columnwidth}
    \centering
    \includegraphics[width=\columnwidth]{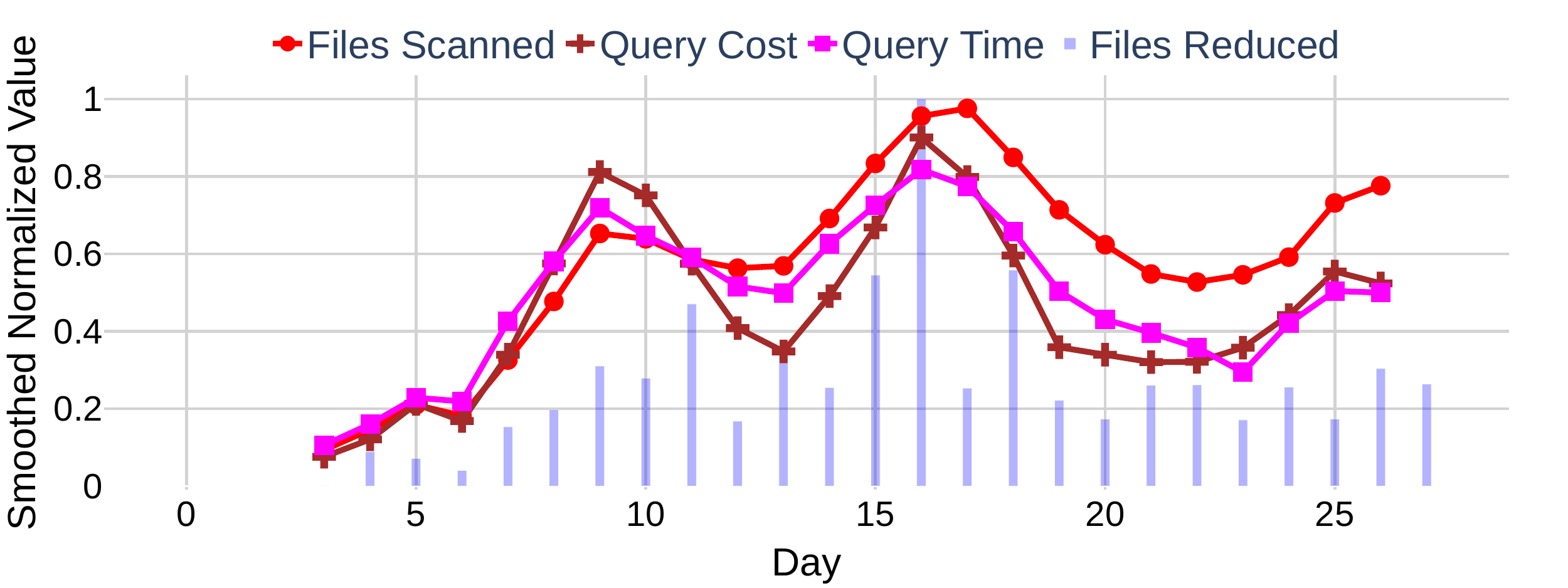}
    \caption{Key workload metrics.}
    \label{fig:pract:workload:files_scanned_post_compaction}
  \end{subfigure}

  \vspace{1em}

  \begin{subfigure}{0.95\columnwidth}
    \centering
    \includegraphics[width=\columnwidth]{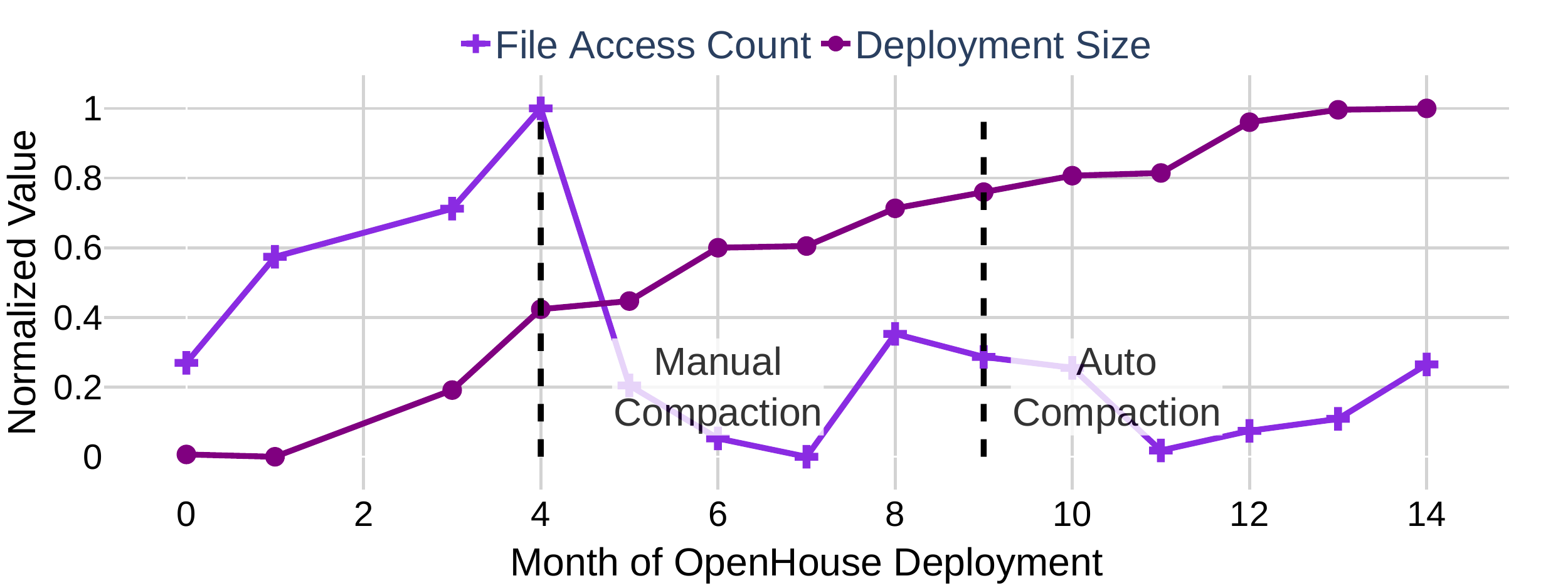}
    \caption{HDFS filesystem operations}
    \label{fig:pract:workload:hdfs_total_open_count}
  \end{subfigure}
  \caption{Impact of \framework on workload metrics, including file scanning, query execution, and HDFS file opens.}
  \label{fig:pract:workload}
\end{figure}

\smallsection{Impact of Compaction on Workloads and HDFS}
We also analyzed the performance of a scan-heavy workload that runs daily, in conjunction with \framework.
To assess the impact of periodic auto-compaction on this workload, we plot the number of files scanned during workload execution and correlated it with query execution time and query cost, as shown in~\Cref{fig:pract:workload:files_scanned_post_compaction}.
The chart captures normalized observations across \(1291\) unique tables chosen by \framework for compaction over the most recent \(30\)-day window.

We observe a strong correlation between compaction runs that reduce file counts and subsequent decreases in files scanned during query execution.
Furthermore, the reduction in files scanned closely corresponds to a decrease in query execution time and query cost, measured in App TBHr.
However, when tables were not selected by \framework for compaction in a given cycle, small files accumulated again--resulting in a recurring sawtooth pattern.

Overall, we observe that despite the increasing size of our deployment over time, the introduction of manual compaction in month \( 4 \) and auto-compaction in month \( 9 \) resulted in a significant reduction in filesystem open() calls on HDFS as shown in ~\Cref{fig:pract:workload:hdfs_total_open_count}.
The sharp decline observed in month \(4\) coincides with the introduction of manual compaction on a subset of heavily fragmented tables, each comprising an average of \(42\text{M}\) small files with an average size of \(64\,\text{MB}\). 
The queries against these tables had previously exhibited frequent HDFS read timeouts due to excessive RPC traffic to the NameNode. 
Such timeouts often led to simultaneous client retries, exacerbating the load and triggering a thundering herd problem.

\section{Discussion and Future Directions}
\label{sec:future}

The lightweight design of \lst implementations such as Apache Iceberg, combined with the deployment of control planes like OpenHouse in data lake-centric architectures, provides an effective setup for defining and implementing data reorganization strategies. 
Through our exploration of the auto-compaction problem both in practice and using our experimental setup, we have identified several exciting directions for future research and innovation in this space.

\smallsection{Navigating Multi-Objective Trade-offs}
In this work, we approached the file compaction problem as a multi-objective optimization task with two primary goals: maximizing file count reduction and minimizing compute cost during compaction. 
Our current method computes a single solution by applying weighted objectives, where the weights reflect our chosen priorities. 
This single solution represents a specific trade-off, providing what we consider to be the \emph{best} answer given the current objective weights and constraints. 
While this approach has shown good results in practice, it inherently risks overemphasizing one metric at the expense of the other by collapsing multiple objectives into a single weighted score. 
In some cases, the chosen solution may not align well with changing system conditions or varying operational requirements, as the optimal balance between file count reduction and compute cost can shift based on workload patterns, resource availability, or specific database needs.
To address these challenges, we propose exploring the use of the Pareto frontier in future work to offer broader perspective on the trade-offs involved. 
Instead of converging on a single \emph{best} answer, leveraging the frontier would allow us to generate a set of Pareto-optimal solutions, each representing a unique balance between file count reduction and compute cost. 
Solutions on the Pareto frontier are non-dominated, meaning that improving one objective would necessarily worsen the other. 
To compute weights dynamically, we propose leveraging regression analysis techniques in machine learning, enabling us to move beyond the reliance on fixed weights for different objectives.

\smallsection{Conflict Resolution}
Our experience revealed that understanding \lst conflict resolution mechanisms and predicting potential conflicts is challenging. 
For example, the experiments with OpenHouse revealed unexpected compaction conflicts involving disjoint partitions, suggesting potential gaps in the \emph{conflict filtering} implementation. 
Recent work, including approaches leveraging formal verification methods, addresses knowledge gaps in conflict resolution for \lsts~\cite{iceberg-cm, delta-cm, hudi-cm}. 
\framework already separates scheduling as an independent step to support diverse \lst conflict resolution strategies, and engine-specific behaviors during maintenance tasks like compaction may still require additional extensions.

\smallsection{Automatic Data Layout Optimization} 
While compaction focuses primarily on managing small files and fragmentation, it can be extended to address broader data layout optimization strategies. 
For instance, data clustering techniques--such as Z-ordering or V-ordering~\cite{delta-zorder,fabric-vorder,hudi-zorder}--can improve compression ratios, encoding efficiency, and query performance by co-locating related data. 
These techniques are complementary to compaction and can be integrated into \framework's decision-making process.
Achieving this integration would require extensions to both the candidate generation phase and the computation of traits. 
For example, some layout optimizations operate at the file level, while others apply at coarser granularities such as partitions or tables. 
These differing scopes would need to be considered during candidate generation. 
Similarly, new traits would need to account for both the benefits of these optimizations--such as improved compression or filtering efficiency--and their costs, including computational overheads like data sampling or multiple data passes. 

\smallsection{Workload Awareness}
Incorporating workload-awareness into trait computation can further refine \framework's decision-making process by aligning layout optimizations with query patterns and access frequency, potentially leading to improvements in query performance. 
Moreover, partitioning and clustering strategies, chosen with query patterns in mind, can also influence the efficiency of writes and compaction by reducing unnecessary data conflict errors~\cite{iceberg-cm}. 
Therefore, the choice of data layout optimization strategy should account not only for query workloads but also for their broader implications on compaction and conflict resolution.

\smallsection{Tuning Write and Compaction Mechanisms and Policies} 
Engines and \lsts expose a wide range of configuration parameters that significantly influence data layout on write. 
For instance, Spark's adaptive query execution framework may inadvertently choose an excessively small shuffle partition size for final writes or a suboptimal distribution mode for table setup, resulting in an excessive number of small files~\cite{DBLP:journals/pvldb/OkolnychyiSTSBHGLT24}. 
In large organizations like LinkedIn, engineers may not have direct control over engine configurations across all workloads. 
Control planes like OpenHouse thus offer a valuable opportunity to analyze and surface such issues, with actionable insights for stakeholders. 
This increased visibility enables timely recommendations to manually mitigate these challenges.
Compaction tasks in \lsts, as well as \framework itself, offer configurable parameters that influence auto-compaction behavior, as discussed in \S\ref{subsec:autotune}. 
Our experiments suggested that ``one size does not fit all'': compaction triggers and data layout strategies should ideally be tailored to individual workloads rather than standardized for all engines. 
However, workload-specific auto-tuning is computationally expensive and in our experiments, each iteration required multiple hours and consumed significant cluster resources. 
Optimizing experimentation time and reducing computational costs will be critical to making auto-tuning feasible for practical use.

\section{Related Work}
\label{sec:related}

The study of automatic compaction has become crucial with the adaptation of DBMS-like structures to \lsts within general-purpose distributed storage systems. 
For example, foundational work by \cite{DBLP:journals/tods/SeveranceL76} introduced \textit{delta files} to mitigate write amplification caused by updates, a concept now central to \lst implementations. 

\smallsection{Database Defragmentation}
\cite{DBLP:conf/cidr/SearsI07} analyzed how object size and data fragmentation affect system performance in the context of a DBMS, highlighting two key findings: First, that optimal data layouts are workload-dependent, and second, that fragmentation over time leads to significant performance degradation. 
DBMS have historically addressed fragmentation using both online and offline approaches. 
Online methods involve human oversight for reorganization. 
For instance, \cite{DBLP:conf/icde/NarasayyaS10} proposed an online approach for index defragmentation in modern databases using a what-if API to estimate performance benefits. 
Their algorithm performs range-level index defragmentation, analogous to fine-grained compaction scopes in \framework. 
The system recommends optimal strategies, but a DBA ultimately schedules and triggers defragmentation. 
In contrast, offline approaches, such as the one adopted by \framework, rely on automated algorithms for reorganization. 
For example, \cite{DBLP:conf/icde/KolovsonS89} described a multi-level index structure that stores mutable data on magnetic disks and immutable archival data on write-once-read-many optical disks. 
A size-based algorithm triggers a vacuum process to migrate data, optimizing read and write latency while reducing storage costs. 
Modern in-memory databases \cite{DBLP:conf/sigmod/DiaconuFILMSVZ13, DBLP:conf/sigmod/SikkaFLCPB12, DBLP:conf/sigmod/ArulrajPM16} apply similar principles by maintaining separate write and read-optimized regions, with data migration triggered by thresholds for size and time. 
Similarly, \cite{DBLP:conf/sigmod/SearsR12} extended these concepts with the bLSM tree, combining B-Tree and LSM tree functionalities. 
Their spring-and-gear scheduler balances compactions across levels, ensuring predictable throughput and consistent latency for uniform workloads. 
Recent work has also systematically explored the LSM compaction design space, analyzing trade-offs across different strategies and workloads~\cite{DBLP:journals/pvldb/SarkarSZA21}. 
\lsts such as Hudi and the recent Apache Paimon~\cite{paimon} incorporate these principles--write and read-optimized regions, along with automatic compaction--directly into their core implementations. 

\smallsection{Automatic Compaction in Data Lakes}
Some engines running on data lakes, while not exposing their table format as \lsts interoperable across engines, employ similar techniques and still produce numerous small files at the storage layer. 
Apache Hive~\cite{DBLP:conf/sigmod/Camacho-Rodriguez19} introduced ACID-compliant tables built on HDFS, employing compaction triggered by thresholds for delta file counts and fragmentation ratios. 
Its design separates cleaning and merging phases to minimize query disruptions, similar to the techniques used in \lsts. 
Nova~\cite{DBLP:conf/sigmod/OlstonCCLHLNRSSTZW11} implemented comparable compaction and cleaning tasks for Apache Pig workflows, but without automation, relying on manual triggers. 
Similarly, \cite{DBLP:journals/pvldb/AhmadK15} proposed a stand-alone compaction server for HBase, isolating compaction tasks from user workloads to improve system efficiency. 
Our \framework deployment, tested in both synthetic experiments and production at LinkedIn, adopts a similar approach. 
However, \framework offers enhanced flexibility, supporting execution in different clusters and multiple operational modes, adapting to diverse workload and system requirements.

\smallsection{Automatic Layout Tuning in \lsts}
Auto-tuning mechanisms have become a critical challenge in \lsts, with recent efforts~\cite{netflix-auto-optimize, aws-iceberg-optimize, azure-databricks-hook, polaris-transactions, hudi-compaction, aws-glue-optimize, aws-s3-tables-optimize} highlighting the need for flexible solutions to address data layout challenges. 
\framework advances these efforts as the first comprehensive proposal for automatic compaction that offers flexibility to adapt to various algorithms, models, and operational scenarios, ensuring compatibility with diverse workloads and infrastructure setups.

\section{Conclusion}
In this paper, we introduced \framework, an automated compaction framework to address small file proliferation, a common problem in data lake infrastructure settings.
\framework is designed according to both functional and non-functional requirements resulting from our deployment at LinkedIn, and its usefulness is shown experimentally with both synthetic and real-world deployments.
We observe that key features such as multi-objective optimization functions and workload-aware compaction strategies and schedules improve the compaction results in our deployments significantly, not only impacting the storage layer positively but also query performance.
We further list a variety of future research opportunities in the broader area of data layout optimization which we think will enhance data lakes effectively moving forward.

\begin{acks}
We would like to thank the entire OpenHouse team at LinkedIn for their collective efforts and contributions. We also extend our heartfelt thanks to Mridul Muralidharan for his thoughtful review of early drafts of this paper and for offering valuable insights.
\end{acks}

\bibliographystyle{ACM-Reference-Format}
\balance
\bibliography{ref}

\end{document}